\documentclass[twocolumn,nofootinbib,prd,aps,amsmath,amssymb,superscriptaddress]{revtex4-1}
\usepackage[varg]{txfonts}
\usepackage{color}
\usepackage{caption}
\usepackage{subcaption}
\usepackage{multirow}
\usepackage{dsfont, mathrsfs}
\usepackage[italicdiff]{physics}
\usepackage{siunitx}
\usepackage{float}
\usepackage{tikz, pgf, pgfplots}
\usepackage{pgfplotstable}
\pgfplotsset{compat=newest}
\usepgfplotslibrary{units} 
\usepackage{hyperref}
\hypersetup{
    colorlinks,
    citecolor=blue,
    filecolor=blue,
    linkcolor=magenta,
    urlcolor=blue
}
\newcommand{\N}{\mathbb N}

\newcommand{\J}{\mathcal{J}}
\newcommand{\I}{\mathrm{I}}

\newcommand{\del}{\partial}
\newcommand\numberthis{\addtocounter{equation}{1}\tag{\theequation}}

\definecolor{MyDarkRed}{rgb}{0.7,0,0}

\newcommand{\edf}[1]{{\textcolor{blue}{#1}}}

\newcommand{\mathsout}[1]
{\bgroup\mathchoice
  {\sbox0{$\displaystyle{#1}$}%
    \usebox0\hspace{-\wd0}%
    \rule[0.5\ht0-0.5\dp0-.5pt]{\wd0}{1pt}}%
  {\sbox0{$\textstyle{#1}$}%
    \usebox0\hspace{-\wd0}%
    \rule[0.5\ht0-0.5\dp0-.5pt]{\wd0}{1pt}}%
  {\sbox0{$\scriptstyle{#1}$}%
    \usebox0\hspace{-\wd0}%
    \rule[0.5\ht0-0.5\dp0-.5pt]{\wd0}{1pt}}%
  {\sbox0{$\scriptscriptstyle{#1}$}%
    \usebox0\hspace{-\wd0}%
    \rule[0.5\ht0-0.5\dp0-.5pt]{\wd0}{1pt}}%
\egroup}

\begin{document}
\title{Nonperturbative collapse models for collisionless self-gravitating flows}
\author{Niels Fardeau}
\email{n.fardeau@ip2i.in2p3.fr}
\author{Thomas Buchert}
\email{buchert@ens--lyon.fr}
\author{Fosca Al Roumi}
\email{fosca.al.roumi@googlemail.com}
\affiliation{Univ Lyon, Ens de Lyon, Univ Lyon1, CNRS, Centre de Recherche Astrophysique de Lyon UMR5574, F-69007, Lyon, France.}
\author{Fereshteh Felegary}
\email{f.felegary@shahroodut.ac.ir}
\affiliation{Faculty of Physics, Shahrood University of Technology, P.O. Box 3619995161 Shahrood, Iran.}

\begin{abstract}
Structure formation in the Universe has been well-studied within the Eulerian and Lagrangian perturbation theories, where the latter performs substantially better in comparison with N-body simulations. Standing out is the celebrated Zel'dovich approximation for dust matter. In this work, we recall the description of gravitational noncollisional systems and extend both the Eulerian and Lagrangian approaches by including, possibly anisotropic, velocity dispersion. A simple case with plane symmetry is then studied with an exact, nonperturbative approach, and various approximations of the derived model are then compared numerically. A striking result is that linearized Lagrangian solutions outperform models based on Burgers' equation in the multistream regime in comparison with the exact solution. These results are finally extended to a 3D case without symmetries, and master equations for the evolution of all parts of the perturbations are derived. The particular 3D case studied corresponds to a maximally anisotropic collapse, which involves an approximation based on the estimation of importance of the different levels of spatial derivatives of the local deformation field.
\end{abstract}

%
\maketitle

\section{Introduction}
\label{intro}
Structures in the Universe, such as galaxies and their clustering into superunits, originate in the gravitational collapse of large, dilute matter clouds. These collapsing clouds are 
well-described by a fluid of massive particles in gravitational interaction but with no collisions with each other, a so-called  collisionless self-gravitating flow, treated in terms of deviations from the homogeneous expansion of the Universe,  described by the commonly adopted Friedmann-Lema\^\i tre-Robertson-Walker (FLRW) models. Gravitational interactions in these systems are to date mostly described in Newtonian theory \cite{bonnor_jeans_1957, peebles:book},  although first studies of this kind were done in the general-relativistic (GR) framework \cite{lifshitz46, lifshitzkhalatnikov}. The dynamics of the collapse is described by the laws of hydrodynamics, which can be derived from conservation laws in Hamiltonian phase space through various approaches, including continuum \textit{mean-field} approximation (Vlasov equation) or the Klimontovic \textit{coarse-graining} approach \cite{binney_galactic_2008, buchert_adhesive_2005}. However, these equations are highly nonlinear and do not admit analytical solutions in the general case. First attempts at describing structure formation in the late Universe used linearized versions of perturbation theory, referred to as the \textit{Eulerian perturbation theory}. This theory is based on the assumption that the density contrast (i.e., the dimensionless deviation from the mean density of the universe model) is small. However, although the volume-averaged density contrast over large regions of the Universe (typically of radius $\sim {10}{\rm Mpc}$) is of order unity, it can reach much larger values in galaxy clusters ($\sim 10$) or in galaxies themselves ($\sim 10^5$). For this reason, the Eulerian perturbation theory is insufficient, and other approaches have to be investigated.\smallbreak
Part of the nonlinearity of the hydrodynamical equations stems from the convective derivative that accounts for the motion of the fluid elements. For this reason, the Eulerian description of the fluid is not best-suited, and a Lagrangian approach is more efficient \cite{serrin}, \cite{ehlersbuchert}, \cite{buchert_varenna}. In the Lagrangian picture, fluid elements themselves constitute the coordinate system, assumed to be constant along the flow, so that there is no longer the need to account for their motion with respect to an Eulerian reference system. Following what was done in Eulerian perturbation theory, new perturbative approaches, based on the Lagrangian framework, appeared. Among them is the celebrated \textit{Zel'dovich approximation} \cite{zeldovich70a, zeldovich70b}, which for the first time described the collapsing elements as highly anisotropic structures (\textit{pancakes}),
rather than spherically symmetric structures as it was understood before. These theories provide solutions for higher-order approximations and encompass nonlinearity to a better extent than Eulerian theories, yielding much more accurate results, even competing with those of many-body simulations.
For a summary and a comprehensive list of references, see the historical account on the Lagrangian approach in Newtonian cosmology in the otherwise GR-based review \cite{Universe}.\footnote{Zel'dovich's approximation has then been transposed to the GR framework (the GR form of Zel'dovich's approximation will not be treated in this work, see \cite{rza1} and subsequent papers in the series for a derivation), where perturbations are no longer described within the global vector space of a Galilei-Newton spacetime, but are intrinsic perturbations of spatial coframe fields in the local rest frames of the fluid.}\smallbreak
All the models mentioned above admit analytical expressions for the evolution of the density contrast in the most simple systems, called \textit{dust fluids}, in which the interaction between fluid elements is only described by gravitation. However, dust fluids have their shortcomings in describing structure formation: as there is no other interaction than gravity between fluid elements, stable structures have a hard time to form in such a system, since they can only be supported by vorticity. The need for a counteracting force led to the development of models including \textit{velocity dispersion} \cite{shukurov}, which were shown to create vorticity in the flow \cite{doroshkevich:vorticity}. In the simplest case velocity dispersion was modeled as an isotropic pressure that allowed stabilizing the structures arising from the collapse. Elliptic galaxies are examples of velocity dispersion-supported structures. Eulerian and Lagrangian perturbative theories were adapted to these new models, along with a phenomenological extension of Zel'dovich's approximation, aiming at describing the system after \textit{shell-crossing singularities} first appeared---coined the ``adhesion approximation'' \cite{gurbatov89} that has been derived from kinetic theory in \cite{buchert:adhesive}, reviewed in \cite{buchert_adhesive_2005} with an explicit coarse-graining method that includes deviations from mean-field gravity. 
Lagrangian methods have been developed to access this regime through effective pressure forces \cite{adler_lagrangian_1998,nonperturbative}, also emphasizing the emergence of vorticity and the impact of anisotropic stresses on the matter power spectrum \cite{ruth:vorticity}. 
Numerical studies started to be performed to check existing models' predictions and explore nonanaly\-tical solutions \cite{klypinshandarin, melottshandarin89}; for recent papers see \cite{rampffrisch,rampf:shellcrossing1,rampf:shellcrossing2,colombi:shellcrossing1,colombi:shellcrossing2}. The reader may again consult the summary paper \cite{Universe} for  further references on the investigation of analytical models and simulations. \smallbreak
In this work, we aim at exploring Lagrangian and nonperturbative approaches to collisionless, self-gravitating flows with anisotropic velocity dispersion within Newtonian cosmology. We will first recall the basic equations in Eulerian and Lagrangian coordinates in Sections~\ref{EJN} and \ref{LJN}. In Section~\ref{1D} we will study the case of plane-symmetric perturbations, both analytically and numerically. In particular, the effect of the size and average density of the clouds on the properties of the collapse phase will be investigated using our numerical integration engine. In Section~\ref{3D} we will extend our reasoning to a more general three-dimensional case without symmetries. In particular, we will be interested in the performance of a Lagrangian linearization of multistream forces in comparison with exact integration of the equations and with models based on Burgers' equation. We conclude in Section~\ref{conclusions}, and we dedicate Appendices to the presentation of proofs and details on the analytical and numerical methods.
\section{The Euler-Jeans-Newton system}
\label{EJN}
To set notations, we use Einstein's summation convention for repeated indices regardless of their (up or down) position (the Newtonian metric is Euclidean). Derivatives with respect to Eulerian coordinates $\{x_i\}$ are denoted by a subscript comma, and derivatives with respect to Lagrangian coordinates $\{X_i\}$ are denoted by a subscript vertical bar, e.g.
\begin{equation*}
    \pdv{f}{x_i}:=\del_if:=f_{,i}\ \qc \pdv{f}{X_i}:=f_{|i} .
\end{equation*}
Vector differential operators without a subscript indicate differentiation with respect to Eulerian coordinates. Vector differential operators with subscript 0 indicate differentiation with respect to Lagrangian coordinates, e.g.
\begin{equation*}
    \boldsymbol{\nabla}\cdot \vb A=A_{k,k}\ \qc \boldsymbol{\nabla}_0 \cdot \vb A=A_{k|k} .
\end{equation*}
Other coordinate systems will be explicitly denoted by an appropriate subscript.
\subsection{General formulation}
We consider a fluid embedded into Galilei-Newton spacetime described by its Hamiltonian phase space density $f(t,\vb x,\vb v)$ in Eulerian phase space coordinates for positions and velocities. Under the hypothesis that the acceleration $\vb b$ is given by the gravitational field $\vb g$ and is therefore velocity-independent, the evolution equation for the phase space density $f$ is given by the Vlasov equation (see e.g. \cite{binney_galactic_2008, buchert:adhesive}),
\begin{equation}
    \pdv{f}{t}+v_i\pdv{f}{x_i}+g_i\pdv{f}{v_i}=0.
\end{equation}
The first moments of this equation provide evolution equations in space for the restmass and momentum density (higher-order velocity tensors are constructed in the same fashion),
\begin{equation}
    \rho :=m\int f\,d^3v\qc \rho\overline{v}_i :=m\int v_if\,d^3v,
\end{equation}
in the form
\begin{subequations}
\begin{gather}
    \pdv{t}\rho+\pdv{x_i}(\rho\overline v_i)=0,\\
    \pdv{t}(\rho\overline v_i)+\pdv{x_j}(\rho\overline{v_iv_j})=\rho g_i.\label{vlasov_1}
\end{gather}
\end{subequations}
Introducing the reduced second velocity moment tensor,
\begin{equation}
    \Pi_{ij}:=\rho\qty(\overline{v_iv_j}-\overline v_i\overline v_j),
\end{equation}
and the Lagrangian time-derivative $d_t=\del_t+\overline v_k\del_k$, we can write (\ref{vlasov_1}) as the \textit{Euler-Jeans} equation
\begin{equation}
    \rho\dv{t}\overline v_i=\rho g_i-\pdv{x_j}\Pi_{ij}.
\end{equation}
The reduced second velocity moment tensor is symmetric and acts like a stress tensor, the divergence of which creates a (multistream) force that counteracts the gravitational field strength. An evolution equation can be found for this tensor, using the second velocity moment of the Vlasov equation (details can be found in Appendix \ref{appA}), in the form
\begin{equation}
    \dv{t}\Pi_{ij}=-\qty[\pdv{\overline v_k}{x_k}\Pi_{ij}+\pdv{\overline v_i}{x_k}\Pi_{jk}+\pdv{\overline v_j}{x_k}\Pi_{ki}]-\pdv{x_k}L_{ijk},\label{evol_pi}
\end{equation}
where $L_{ijk}$ is the third reduced velocity moment 
\begin{equation}
    L_{ijk}:=\rho\overline{(v_i-\overline v_i)(v_j-\overline v_j)(v_k-\overline v_k)}.
\end{equation}
However, this equation cannot yield a closed system, due to the introduction of the third-order tensor $L_{ijk}$. One way to close it is to consider that deviations from the mean velocity are small (which is valid in the pre-virialization regime and before any shell-crossing), typically $|\vb v-\overline{\vb v}|\sim\epsilon|\overline{\vb v}|$, which allows to estimate the higher-order moments $\Pi_{ij}\sim\rho\epsilon^2|\overline{\vb v}|^2$ and $L_{ijk}\sim\rho\epsilon^3|\overline{\vb v}|^3$. If $\epsilon$ is small enough, one can drop the $L_{ijk}$ term in (\ref{evol_pi}). For later stages of evolution, the omission of the third-order term remains phenomenological, which should be kept in mind when solutions of the following system are studied. With this truncation of the hierarchy, the resulting system, along with the Newtonian field equations for the gravitational field strength, is called the \textit{Euler-Jeans-Newton} (EJN) system and reads:
\begin{subequations}\label{ejn}
\begin{gather}
    \dv{t}\rho=-\rho\overline v_{k,k},\label{ejna}\\
    \dv{t}\overline v_i=g_i-\dfrac{1}{\rho}\Pi_{ij,j},\label{ejnb}\\
    \dv{t}\Pi_{ij}=-\qty[\overline v_{k,k}\Pi_{ij}+\overline v_{i,k}\Pi_{jk}+\overline v_{j,k}\Pi_{ki}],\label{ejnc}\\
    g_{[i,j]}=0\qc g_{k,k}=\Lambda-4\pi G\rho.\label{ejnd}
\end{gather}
\end{subequations}
\subsection{Hubble-comoving frame and deviation fields}
As it is usually done in cosmology, we introduce Eulerian \textit{Hubble-comoving} coordinates $\vb q=\vb x/a(t)$ with the isotropic and homogeneous expanding Friedmannian background (with scale factor $a(t)$), and we split the dependent variables as follows \cite{peebles:book}, \cite{buchert_adhesive_2005}:
\begin{subequations}\label{peculiar}
\begin{gather}
    \overline{\vb v}(\vb q,t)={\vb v}_H+\vb u(\vb q,t),\label{peculiara}\\
    \vb g(\vb q,t)=\vb g_H+\vb w(\vb q,t),\\
    \rho(\vb q,t)=\rho_H\qty[1+\delta(\vb q,t)],\\
    \Pi_{ij}(\vb q,t)=\pi^H_{ij}+\pi_{ij}(\vb q,t),
\end{gather}
\end{subequations}
where homogeneous quantities are given in the background by ${\vb v}_H=\dot a\vb q$, $\vb g_H=\ddot a\vb q$ and $\pi_{ij}^H=p_Ha^2\delta_{ij}$.\footnote{We, henceforth, omit the overbar for the homogeneous and the deviation fields with the understanding that both are mean quantities in the fluid description.} The homogeneous density $\rho_H$ is given by the integral of the continuity equation $\rho_H=\rho_{H \rm i}/a^3$ where $\rho_{H \rm i}$ denotes the background density at some initial time. The variables $\vb u,\vb w$ and $\delta$ are usually called peculiar-velocity, 
peculiar-acceleration and density contrast. In Hubble-comoving coordinates, the differential operators transform as 
\begin{equation}
    \grad=\dfrac{1}{a}\grad_{\vb q}\qc\eval{\pdv{t}}_{\vb x}=\eval{\pdv{t}}_{\vb q}-H\vb q\cdot\grad_{\vb q},
\end{equation}
with $H=\dot a/a$ the Hubble function. Within this framework, we can split the EJN system (\ref{ejn}) using (\ref{peculiar}). Assuming Friedmann's differential equations to hold for the (possibly relativistic) background model (in particular, neglecting any backreaction of inhomogeneities on the latter), we obtain the EJN system for the deviation fields:
\begin{subequations}\label{ejn_pec}
\begin{gather}
    \dv{t}\delta+\dfrac{1}{a}(1+\delta)\pdv{u_k}{q_k}=0,\label{ejn_peca}\\
    \dv{t}u_i+Hu_i=w_i-\dfrac{1}{a}\dfrac{1}{\rho_H(1+\delta)}\pdv{q_j}\pi_{ij},\label{ejn_pecb}\\
    \dv{t}\pi_{ij}+5H\pi_{ij}=-\dfrac{1}{a}\qty[\pdv{u_k}{q_k}\pi_{ij}+\pdv{u_i}{q_k}\pi_{jk}+\pdv{u_j}{q_k}\pi_{ki}]
    \nonumber\\
    -p_Ha^2\qty[\qty(7H+\dfrac{1}{a}\pdv{u_k}{q_k})\delta_{ij}+\dfrac1a\qty(\pdv{u_i}{q_j}+\pdv{u_j}{q_i})],
    \label{ejn_pecc}\\
    \pdv{w_i}{q_j}-\pdv{w_j}{q_i}=0\qc \pdv{w_k}{q_k}=-4\pi Ga\rho_H\delta.\label{ejn_pecd}
    \end{gather}
\end{subequations}
\section{Lagrangian formulation of the Euler-Jeans-Newton system}
\label{LJN}
In this section, we base our analysis on the detailed investigation for an isotropic-pressure supported fluid in \cite{adler_lagrangian_1998}, and generalize to nonisotropic velocity dispersion. The reader may also consult the investigation in \cite{ruth:vorticity} within the Lagrangian picture, emphasizing the impact of anisotropic stresses in dispersion-supported fluids.

\subsection{Lagrangian picture}
We consider that the Eulerian coordinates $\vb x$ can be related to the Lagrangian coordinates, $\vb X$, \textit{via} a time-dependent diffeomorphism $\vb f$, that is 
\begin{equation}
    \vb x=\vb f(\vb X,t),
\end{equation}
with $f(\vb X,t_0)=\vb X$. Lagrangian coordinates stay attached to fluid elements along their trajectories, meaning in particular that ${\dv{t}} {\vb X}={\bf 0}$. The Lagrangian time-derivative thus merely reads as the partial time-derivative in the Lagrangian frame,
\begin{equation}
    \dv{t}=\eval{\pdv{t}}_{\vb X}=\eval{\pdv{t}}_{\vb x}+\vb v\cdot\grad.
\end{equation}
We also define the Jacobian matrix of the transformation from Lagrangian to Eulerian coordinates, and its determinant,
\begin{equation}
    J_{ik}(\vb X,t):=\pdv{f_i}{X_k}\qc J(\vb X,t):=\det J_{ik}(\vb X,t).
\end{equation}
According to usual calculus rules, volume elements then transform as
\begin{equation}
    d^3x=Jd^3X.
\end{equation}
In the Lagrangian framework, the diffeomorphism $\vb f$ is the only dynamical variable. The velocity and acceleration fields are directly given by the definitions
\begin{equation}
    \vb v(\vb X,t)=\dot{\vb f}(\vb X,t)\qc\vb b(\vb X,t)=\ddot{\vb f}(\vb X,t).
\end{equation}
The continuity equation, which represents the total restmass conservation within a compact domain $D_t$ (evolving in time in the Eulerian frame, but its boundaries are comoving with the fluid), is solved by an exact integral for the density field as a functional of $\vb f$. Indeed, 
\begin{align*}
    \dv{t}M_{D_t}=0 \nonumber\\
    =\dv{t}\int_{D_t}\rho(\vb x,t)\,d^3x=\dv{t}\int_{D_{t_0}}\rho(\vb X,t)J(\vb X,t)\,d^3X\\
    =\int_{D_t}\dfrac{1}{J}\dv{t}\qty(\rho(\vb X,t)J(\vb X,t))\,d^3x,\numberthis{}
\end{align*}
which implies that the quantity $\rho J$ is conserved along flow lines, i.e.
\begin{equation}
    \rho J=\rho_{\rm i}J_{\rm i}:=C(\vb X)\Rightarrow \rho(\vb X,t)=\dfrac{C(\vb X)}{J(\vb X,t)}.\label{rho_lagr}
\end{equation}
(Since $J_{\rm i} = 1$, $C = \rho_{\rm i}$.) 
\subsection{Lagrangian derivative}
In the process of getting rid of Eulerian coordinates, we need to transfer Eulerian derivatives to the Lagrangian frame \cite{adler_lagrangian_1998}. To do so, we consider the inverse diffeomorphism of $\vb f$,
\begin{equation}
    \vb h :=\vb f^{-1}\qc\text{i.e.}\quad \vb X=\vb h(\vb x,t).
\end{equation}
Using this definition, we can write
\begin{subequations}
\begin{equation}
    \pdv{x_i}=\pdv{h_k}{x_i}\pdv{X_k}=J^{-1}_{ik}\pdv{X_k},
\end{equation}
and with the definition of the adjoint matrix,
\begin{equation}
\mathrm{ad}\, J_{ik}:=\frac{1}{2}\epsilon_{klm}\epsilon_{ipq}f_{p|l}f_{q|m},   
\end{equation}
we get the transformation rule
\begin{equation}\label{euler_dv_full}
    \pdv{x_i}=\dfrac{1}{2J}\epsilon_{klm}\epsilon_{ipq}f_{p|l}f_{q|m}\pdv{X_k}.
\end{equation}
\end{subequations}
To further lighten the notations, we define the functional determinant 
\begin{subequations}
\begin{equation}
    \J(A,B,C)=\dfrac{\del(A,B,C)}{\del(X_1,X_2,X_3)}=\epsilon_{ikl}A_{|i}B_{|k}C_{|l}.
\end{equation}
The functional determinant satisfies the usual rules of determinants, plus the Leibniz rule,
\begin{equation}
    \J(AD,B,C)=A\J(D,B,C)+D\J(A,B,C).
\end{equation}
\end{subequations}
Using this notation and Eq. (\ref{euler_dv_full}), the Eulerian derivative of a field $\cal A$ can be written as
\begin{equation}\label{lagr_tsfo}
    {\cal A}_{,i}=\dfrac{1}{2J}\epsilon_{ikl}\J({\cal A},f_k,f_l).
\end{equation}
\subsection{Lagrangian form of the field equations}
As it has been said above, the continuity equation and the momentum conservation equation are automatically solved in the Lagrangian picture through definitions and an exact integral,
\begin{equation}\label{lagr_int}
    \vb v=\dot{\vb f}\qc \vb b=\ddot{\vb f}\qc \rho=\dfrac{\rho_0J_0}{J}.
\end{equation}
Let us now consider the field equations of the Euler-Jeans-Newton system. With (\ref{ejnb}), we can first write
\begin{equation}
    \vb g=\vb b+\dfrac{1}{\rho}\boldsymbol{\psi}\qc \psi_i \equiv \Pi_{ij,j}=\pi_{ij,j}.
\end{equation}
Using vector identities, Equation (\ref{ejnd}) takes the form
\begin{subequations}\label{field_vec}
\begin{gather}
    \boldsymbol{\nabla}\times \vb b+\dfrac{1}{\rho}\boldsymbol{\nabla}\times \boldsymbol{\psi}-\dfrac{1}{\rho^2}\grad\rho\times\boldsymbol{\psi}=\vb 0,\\
    \boldsymbol{\nabla}\cdot \vb b+\dfrac{1}{\rho}\boldsymbol{\nabla}\cdot\boldsymbol{\psi}-\dfrac{1}{\rho^2}\boldsymbol{\psi}\cdot\grad\rho=\Lambda-4\pi G(\rho+3p_H).
\end{gather}
\end{subequations}
As we may consider a relativistic Friedmannian background, the homogeneous pressure $p_H$ appears in the acceleration law and as a source of the gravitational field (in general relativity, a pressure is self-gravitating, whereas it is not in Newtonian cosmology). We then transform the Eulerian derivatives as in (\ref{lagr_tsfo}), and we recall the identity $\epsilon_{jik}\epsilon_{jlm}=\delta_{il}\delta_{km}-\delta_{im}\delta_{kl}$, to obtain the \textit{Lagrange-Jeans-Newton system} (LJN):
\begin{subequations}
\begin{gather}
    \dfrac{1}{J}\J(\ddot f_k,f_k,f_i)+\dfrac{1}{\rho J}\J(\psi_k,f_k,f_i)-\dfrac{1}{\rho^2J}\psi_k\J(\rho,f_k,f_i)=0,\\
    \dfrac{1}{2J}\epsilon_{kij}\J(\ddot f_k,f_i,f_j)+\dfrac{1}{2\rho J}\epsilon_{kij}\J(\psi_k,f_i,f_j)\nonumber\\
    -\dfrac{1}{2\rho^2J}\epsilon_{kij}\psi_k\J(\rho,f_i,f_j)=\Lambda-4\pi G(\rho+3p_H).
\end{gather}
\end{subequations}
Replacing $\psi_k$ by its Lagrangian expression $\psi_k=\pi_{kp,p}=\dfrac{1}{2J}\epsilon_{pqr}\J(\pi_{kp},f_q,f_r)$, this becomes
\begin{subequations}\label{lagr_raw}
\begin{eqnarray}
    &\dfrac{1}{J}\J(\ddot f_k,f_k,f_i)
    +\dfrac{1}{2\rho J}\epsilon_{pqr}\J\qty(\dfrac{1}{J}\J(\pi_{kp},f_q,f_r),f_k,f_i)\nonumber\\
    &-\dfrac{1}{2\rho^2J^2}\epsilon_{pqr}\J(\rho,f_k,f_i)\J(\pi_{kp},f_q,f_r)=0,
\end{eqnarray}\vspace{-.55cm}
\begin{eqnarray}
    &\dfrac{1}{2J}\epsilon_{kij}\J(\ddot f_k,f_i,f_j)\nonumber\\
    &+\dfrac{1}{4\rho J}\epsilon_{kij}\epsilon_{pqr}\J\qty(\dfrac{1}{J}\J(\pi_{kp},f_q,f_r),f_i,f_j)\nonumber\\
    &-\dfrac{1}{4\rho^2J^2}\epsilon_{kij}\epsilon_{pqr}\J(\rho,f_i,f_j)\J(\pi_{kp},f_q,f_r)\nonumber\\
    &=\Lambda-4\pi G(\rho+3p_H).
\end{eqnarray}
\end{subequations}
We now introduce a common approximation\footnote{This approximation can be formally avoided through an exact argument that can be found in appendix A of \cite{adler_lagrangian_1998}.} in cosmology: writing $\rho(\vb X,t)=\rho_H(t)\qty(1+\delta(\vb X,t))$, we consider that at the initial time $t_{\rm i}$, the density contrast is small, so that the density is approximately homogeneous $\rho(\vb X,t_{\rm i})\approx \rho_H(t_{\rm i})$. This approximation is particularly good at the time of recombination (or time of decoupling, $t_{\rm i}=t_{\rm rec}=\sim {3.8}\times 10^5\;{y}$). Therefore, we have $C_{|i}:={C_{H}}_{|i}=0$ that simplifies the equations substantially. We use the notation $C_H$ to remind the reader of this step, $C_H=\rho_H J=\rho_H(t_{\rm i})$. Using this approximation and the integral (\ref{lagr_int}) for the density, the LJN system \eqref{lagr_raw} simplifies to 
\begin{subequations}\label{lagr_homog}
\begin{gather}
    \dfrac{1}{J}\J(\ddot f_k,f_k,f_i)+\dfrac{1}{2C_HJ}\epsilon_{pqr}\J(\J(\pi_{kp},f_q,f_r),f_k,f_i)=0,\\
    \dfrac{1}{2J}\epsilon_{kij}\J(\ddot f_k,f_i,f_j)+\dfrac{1}{4C_HJ}\epsilon_{kij}\epsilon_{pqr}\J(\J(\pi_{kp},f_q,f_r),f_i,f_j)\nonumber\\
    =\Lambda-4\pi G(\rho+3p_H).
\end{gather}
\end{subequations}
Finally, we decompose $\pi_{ij}=p\delta_{ij}+\Delta_{ij}$ into an isotropic kinetic pressure function and a traceless part, and we make the further assumption that the isotropic pressure is also barotropic, that is $p=\alpha(\rho)$. Using the antisymmetry of the Levi-Civita tensor and of the functional determinant, one can show that the LJN system takes the following form (where $\alpha'$ indicates the derivative $d\alpha/d\rho$):
\begin{subequations}\label{lagr_p}
\begin{eqnarray}
    &\dfrac{1}{J}\J(\ddot f_k,f_k,f_i)\nonumber\\
    &+\dfrac{1}{2C_HJ}\epsilon_{pqr}\J(\J(\Delta_{kp},f_q,f_r),f_k,f_i)=0,\label{lagr_pa}\\
    &\dfrac{1}{2J}\epsilon_{kij}\J(\ddot f_k,f_i,f_j)\nonumber\\
    &+\qty(\dfrac{C_H\alpha''}{2J^5}+\dfrac{\alpha'}{J^4})\J(J,f_i,f_j)\J(J,f_i,f_j)\nonumber\\
    &-\dfrac{\alpha'}{2J^3}\J(\J(J,f_i,f_j),f_i,f_j)
    \nonumber\\
    &+\dfrac{1}{4C_HJ}\epsilon_{kij}\epsilon_{pqr}\J(\J(\Delta_{kp},f_q,f_r),f_i,f_j)\nonumber\\
    &=\Lambda-4\pi G\qty(\dfrac{C_H}{J}+3p_H).\label{lagr_pb}
\end{eqnarray}
\end{subequations}
\subsection{Principal scalar invariants}
We now wish to define the useful notion of principal scalar invariants of a tensor. Given an order 2 tensor 
$\mathbb{A}=(A_{ij})$ in three dimensions, with eigenvalues $\lambda_1,\lambda_2,\lambda_3$, the following quantities are independent of the frame in which the tensor is written \cite{bks}:
\begin{subequations}
\begin{gather}
    \mathrm I(\mathbb A)=\Tr(\mathbb A)=A_{ii}=\lambda_1+\lambda_2+\lambda_3,\\
    \mathrm{II}(\mathbb A)=\dfrac12\qty(\Tr(\mathbb A)^2-\Tr\big(\mathbb A^2))=\dfrac12\qty(A_{ii}A_{jj}-A_{ij}A_{ji})\nonumber\\
    =\lambda_1\lambda_2+\lambda_2\lambda_3+\lambda_1\lambda_3,\\
    \mathrm{III}(\mathbb{A})=\det\mathbb{A} =\dfrac{1}{6}\epsilon_{ijk}\epsilon_{lmn}A_{il}A_{jm}A_{kn}=\lambda_1\lambda_2\lambda_3.
\end{gather}
\end{subequations}
In particular, we are interested in the invariants of the gradient of the deformation field $\mathbf {d P}$ describing the deviation from the homogeneous and isotropic background expansion. We can write
\begin{equation}\label{ansatz_gen}
    \vb f(\vb X,t)=a(t)[\vb X+\vb P(\vb X,t)],
\end{equation}
with $\vb P(\vb X,t)$ the scaled deviation from the background, and we define $\mathrm{I}=\mathrm{I}(\grad_0\vb P), \mathrm{II}=\mathrm{II}(\grad_0\vb P), \mathrm{III}=\mathrm{III}(\grad_0\vb P)$, where $\grad_0$ indicates the gradient with respect to the Lagrangian coordinates, with components $P_{i|j}$. Using the above definitions and the properties of the Levi-Civita tensor, one can show the following identities \cite{bks}:
\begin{eqnarray}
    &\epsilon_{ijk}\J(X_i,X_j,X_k)=6\qc \epsilon_{ijk}\J(X_i,X_j,P_k)=2\,\mathrm{I}\nonumber\\
    &\epsilon_{ijk}\J(X_i,P_j,P_k)=2\,\mathrm{II}\qc
    \epsilon_{ijk}\J(P_i,P_j,P_k)=6\,\mathrm{III},\nonumber\\
\end{eqnarray}
which in turn yield an expression for the Jacobian determinant of the Lagrangian transformation 
\begin{equation}
    J=\det\qty(\grad_0\vb f)=a^3\qty(1+\mathrm{I}+\mathrm{II}+\mathrm{III}).
\end{equation}
\section{Nonperturbative plane-symmetric gravitational collapse with velocity dispersion}
\label{1D}
\subsection{Motivation}\label{par:motiv}
The first two sections of this work gave the theoretical framework needed to investigate large-scale structure formation in the late Universe. In the following, we will apply this framework to systems with isotropic (plane-symmetric) and anisotropic (three-dimensional) velocity dispersion, with a nonperturbative approach. \smallbreak
Before diving into those cases, it is useful to remind the results obtained from standard Eulerian perturbation theory (see e.g. \cite{peebles:book}, for a systematic list of the different approximations involved and the notion of ``nonperturbative'', see \cite{nonperturbative}). Starting from the Euler-Jeans-Newton system in the Eulerian comoving picture (\ref{ejn_pec}), considering only isotropic pressure with a dynamical equation of state $\pi_{ij}=p\delta_{ij}=\alpha(\rho)\delta_{ij}$, and keeping only terms of order 1 in peculiar-fields, one obtains the following Eulerian evolution equation for the density contrast $\delta$:
\begin{equation}
    \eval{\pdv[2]{t}}_{\vb q}\delta+2H\eval{\pdv{t}}_{\vb q}\delta-4\pi G\rho_H\delta=\dfrac{\alpha'}{a^2}\laplacian_{\vb q}\delta.
\end{equation}
The right-hand side of this equation may not be linear due to the term $\alpha'(\rho)$, which can have any dependence on $\rho$ and shall then be linearized too. Here we can also introduce a characteristic length for self-gravitating systems with pressure, called \textit{Jeans' length}, and reading
\begin{equation}
    L_J^2=\dfrac{\alpha'}{4\pi G\rho_H},
\end{equation}
so that the above equation can be rewritten as
\begin{equation}
    \eval{\pdv[2]{t}}_{\vb q}\delta+2H\eval{\pdv{t}}_{\vb q}\delta-4\pi G\rho_H\delta=4\pi G\rho_H\dfrac{L_J^2}{a^2}\laplacian_{\vb q}\delta.
\end{equation}
Jeans' length represents the typical size of the self-gravitating system below which the system collapses under its own gravity, and above which it expands (for more details on Eulerian perturbation theory and self-gravitating systems stability, see \cite{binney_galactic_2008}).
\smallbreak
Moving now to the LJN system (\ref{lagr_p}) with isotropic pressure, using the ansatz (\ref{ansatz_gen}) and linearizing with respect to $\vb P$, one obtains a similar evolution equation for the longitudinal (irrotational) part $\vb P^L$ of $\vb P$:
\begin{equation}\label{lagr_lin}
    \ddot{\vb P}^L+2H\dot{\vb P}^L-4\pi G\rho_H\vb P^L=\dfrac{\alpha'}{a^2}\laplacian_0\vb P^L,
\end{equation}
which can be written in terms of $\delta$ using the (linearized) relation $\delta=-\grad_0\cdot\vb P=-\grad_0\cdot\vb P^L$. However, it is to be noted that Lagrange-linear equations are not Euler-linear, due to the presence of the convective derivative. The Lagrangian approach is \textit{intrinsically nonlinear}. Yet, although Lagrange-linear approximation allows us to encompass nonlinearity to some extent, it is still a perturbative approach and relies on various approximations, which can be found in detail in \cite{nonperturbative}, 
along with the most general evolution equation for the perturbations. \smallbreak
In the following, we will first solve the LJN system (\ref{lagr_p}) in the case of plane symmetry \textit{exactly} using Burgers' equation and Fourier analysis, which removes the anisotropy of velocity dispersion and renders the equations more manageable. Then, we will move on to three-dimensional anisotropic deviations, which we will solve by using a ``locally one-dimensional'' approach arising from a maximal anisotropy hypothesis.  
\subsection{Plane-symmetric collapse}
In the case of plane symmetry, a solution for the Lagrange-linearized version of (\ref{lagr_p}) has been found in \cite{al_roumi_matiere_2011}.
Here we will take the same path to obtain the evolution equation for the deviation field (in Newtonian theory), but we will then explore a different approach to find solutions to the latter, which will be useful to solve the 3D case. Note that in the case of an Einstein-de Sitter background, exact solutions in the form of power series have been made explicit in \cite{rampffrisch, rampf:shellcrossing1, rampf:shellcrossing2}. \smallbreak
We start with the following ansatz for the deformation field: 
\begin{equation}\label{ansatz_1d}
    f_1=a(X_1+P_1(X_1,t))\qc f_2=aX_2\qc f_3=aX_3,
\end{equation}
where the background is expanding in all three dimensions, but the deviations from isotropic expansion only happen in one direction. The goal here is to find a closed equation for the evolution of the deviation $P_1$. Using our ansatz, we can calculate the different terms of Equation (\ref{lagr_pb}):\footnote{Equation (\ref{lagr_pa}) is trivial with this ansatz.}
\begin{subequations}
\begin{gather}
    J=a^3(1+P_{1|1}),\\
    \J(J,f_i,f_j)=a^5\epsilon_{1ij}P_{1|11},\\
    \J(\J(J,f_i,f_j),f_i,f_j)=2a^7P_{1|111},\\
    \J(\Delta_{kp},f_q,f_r)=a^2\epsilon_{1qr}\Delta_{kp|1},\\
    \epsilon_{kij}\epsilon_{pqr}\J(\J(\Delta_{kp},f_q,f_r),f_i,f_j)=4a^4\Delta_{11|11},\\
    \epsilon_{kij}\J(\ddot f_k,f_i,f_j)=6\ddot aa^2(1+P_{1|1})+4\dot aa^2\dot P_{1|1}+2a^3\ddot{P}_{1|1},
\end{gather}
\end{subequations}
which finally lead to the equation
\begin{eqnarray}
  &3\dfrac{\ddot a}{a}+2\dfrac{\dot a}{a(1+P_{1|1})}\dot P_{1|1}+\dfrac{\ddot P_{1|1}}{1+P_{1|1}}\nonumber\\
  &+\qty(\dfrac{C_H\alpha''}{a^5(1+P_{1|1})^5}+\dfrac{2\alpha'}{a^2(1+P_{1|1})^4})P_{1|11}^2 \nonumber\\
  &-\dfrac{\alpha'}{a^2(1+P_{1|1})^3}P_{1|111}
  +\dfrac{a}{C_H(1+P_{1|1})}\Delta_{11|11}\nonumber\\
  &=\Lambda-4\pi G\qty(\dfrac{C_H}{a^3(1+P_{1|1})}+3p_H).
\end{eqnarray}
Injecting Friedmann's acceleration law for the homogeneous background,
\begin{equation}\label{fried}
    3\dfrac{\ddot a}{a}=\Lambda-4\pi G\qty(\dfrac{C_H}{a^3}+3p_H),
\end{equation}
we are left with
\begin{eqnarray}
  &2\dfrac{\dot a}{a}\dot P_{1|1}+\ddot P_{1|1}+\qty(\dfrac{C_H\alpha''}{a^5(1+P_{1|1})^4}+\dfrac{2\alpha'}{a^2(1+P_{1|1})^3})P_{1|11}^2 \nonumber\\
  &-\dfrac{\alpha'}{a^2(1+P_{1|1})^2}P_{1|111}
  +\dfrac{a}{C_H}\Delta_{11|11}\nonumber\\
  &=4\pi G\dfrac{C_H}{a^3}P_{1|1},
\end{eqnarray}
which can conveniently be rewritten as a derivative,
\begin{eqnarray}\label{p1_dv}
    &\qty(\ddot P_1+2H\dot P_1-\dfrac{4\pi GC_H}{a^3}P_1)_{|1}\nonumber\\
    &=\qty(\dfrac{\alpha'P_{1|11}}{a^2(1+P_{1|1})^2}-\dfrac{a}{C_H}\Delta_{11|1})_{|1}.
\end{eqnarray}
Relevant initial conditions are prescribed, following the notations developed in the relativistic context in \cite{rza3, rza4}, 
\begin{equation}
    \mathscr{P}_1=0\qc \dot{\mathscr P}_1=U_1\qc \ddot{\mathscr P}_1=W_1-2HU_1,
\end{equation}
where $\mathscr P_i, U_i$ and $W_i$ are shorthand notations for the deviation fields, peculiar-velocity and -acceleration at the initial time $t_{\rm i}$. However, within the homogeneous approximation ${C_H}_{|i}=0$ used previously, there is no initial peculiar-acceleration (see the field equations (\ref{ejn_pecd}) with $\delta\approx0$), so that we can take $W_i\approx0$. Using these initial conditions, the integration constant arising when integrating (\ref{p1_dv}) can be set to zero, leaving us with
\begin{equation}\label{p1_delta}
    \ddot P_1+2H\dot P_1-\dfrac{4\pi GC_H}{a^3}P_1=\dfrac{\alpha'P_{1|11}}{a^2(1+P_{1|1})^2}-\dfrac{a}{C_H}\Delta_{11|1}.
\end{equation}
Linearizing this equation with respect to $P_1$ and neglecting anisotropy $\Delta_{11}$, we find an equation similar to (\ref{lagr_lin}). Now, pursuing our nonperturbative approach, we need to close the previous equation. To do so, we have to relate the velocity dispersion to other variables of the system. As the velocity field is homogeneous in the directions $X_2$ and $X_3$, the only nonvanishing component of the peculiar-velocity dispersion tensor $\pi_{ij}$ is $\pi_{11}$, so that we can write:
\begin{eqnarray}
    &\pi_{ij}=\begin{pmatrix}\pi_{11}&0&0\\0&0&0\\0&0&0\end{pmatrix}=\underbrace{\begin{pmatrix}\frac{1}{3}\pi_{11}&0&0\\0&\frac{1}{3}\pi_{11}&0\\0&0&\frac{1}{3}\pi_{11}\end{pmatrix}}_{p\delta_{ij}}\nonumber\\
    &+\underbrace{\begin{pmatrix}\frac{2}{3}\pi_{11}&0&0\\0&-\frac{1}{3}\pi_{11}&0\\0&0&-\frac{1}{3}\pi_{11}\end{pmatrix}}_{\Delta_{ij}}.
\end{eqnarray}
The evolution equation for $\pi_{ij}$ (\ref{ejn_pecc}) reads for the only component $\pi_{11}$:
\begin{equation}\label{pi11_dot_p}
    \pdv{\pi_{11}}{t}+\dfrac{u_1}{a}\pdv{\pi_{11}}{q_1}+5H\pi_{11}
    =-\dfrac{1}{a}\pdv{u_1}{q_1}\,\!(3\pi_{11}+p_Ha^2)-7p_Ha\dot a.
\end{equation}
After the time of recombination, Friedmann's expansion law starts to be dominated by matter, and we may neglect the background pressure $p_H$ (being, however, nonvanishing also in the matter-dominated regime) leaving only
\begin{equation}\label{pi11_dot}
    \pdv{\pi_{11}}{t}+\dfrac{u_1}{a}\pdv{\pi_{11}}{q_1}+5H\pi_{11}=-\dfrac{3}{a}\pdv{u_1}{q_1}\pi_{11}.
\end{equation}
Solutions of this equation are polytropic with adiabatic exponent $\gamma=3$ (see Appendix \ref{appA} for proof): 
\begin{equation}\label{pi11}
    \pi_{11}=\beta a^4\rho^3,
\end{equation}
where $\beta$ can be chosen constant with appropriate initial conditions. Knowing this, the right-hand side of Equation (\ref{p1_delta}) simplifies to
\begin{equation}
    \label{p1_beta}
    \ddot P_1+2H\dot P_1-\dfrac{4\pi GC_H}{a^3}P_1=-\dfrac{3\beta C_H^2}{a^4(1+P_{1|1})^4}P_{1|11}.
\end{equation}
\subsection{Explicit solution}
Equation (\ref{p1_beta}) has been solved in \cite{al_roumi_matiere_2011} after linearization in $P_1$ using Fourier analysis, and the following expression was found:
\begin{equation}
    P_1(X_1,t)=\int_{-\infty}^\infty C_1(X_1-S)A(S,t)dS,
\end{equation}
where
\begin{eqnarray}
    &A(S,t)=\nonumber\\
    &\lim\limits_{y\to0^+}\qty(\dfrac{a}{2y}+2\Re\qty[\mathcal{G}(S)\int_{\;\qty(\sqrt{\frac{6c}{a}}-iS)\;y}^\infty \dfrac{e^{-z}}{z}\dd z])\nonumber\\ &\text{with} \ \  \mathcal{G}(S)=\dfrac{aS^2+i\sqrt{6ac}S}{2c\qty(\sqrt{\frac{6c}{a}}-iS)}\nonumber \\
    &\text{and} \ \ C_1(X_1)=\int k_1^2 e^{-k_1^2}e^{ik_1X_1}\dd k_1.
\end{eqnarray}
Here, we want to be more general than the linear approximation. In the case of plane symmetry, both the peculiar-acceleration $\vb w$ and the peculiar-velocity $\vb u$ have only one nonzero component in the direction $q_1$, and are therefore parallel, so that we can write in general
\begin{equation}\label{slaving}
    \vb w=h(\vb q,t)\vb u.
\end{equation}
This condition is not generally verified in the absence of symmetry in the model, but is usually postulated as an approximation (so-called ``slaving condition''). Assuming further that the proportionality factor $h$ is the same as in the Eulerian linear \textit{dust} (pressureless) model (and in particular, only depends on time), we obtain the so-called ``adhesion approximation'' \cite{gurbatov89}, \cite{buchert_adhesive_2005, extending} :
\begin{equation}\label{parallel}
    \vb w=h(t)\vb u\ \ ~~\edf{,}~~ h(t)=4\pi G\rho_H\dfrac{b}{\dot b}\ \ ~~\edf{,}~~ \ddot b+2H\dot b-4\pi G\rho_H b=0.
\end{equation}
This approximation is supposed to be valid in the weakly nonlinear regime with pressure, before any shell-crossing singularities appear. Injecting this approximation into the EJN system (\ref{ejn_pec}), along with the expression (\ref{pi11}) for the pressure term, changing the time variable to $b$ and rescaling the peculiar-velocity as $\tilde u=u_1/a\dot b$, we obtain the equation:
\begin{equation}\label{burgers}
    \dv{\tilde u}{b}=\mu(\rho,t)\pdv[2]{\tilde u}{q_1}\qc \mu(\rho,t)=3\beta\rho_{H,0}^2(1+\delta)\dfrac{b}{a^4\dot b^2}.
\end{equation}
The coefficient $\mu$ formally plays the role of a time and density-dependent viscosity term, although the equations are fully time-reversible. If $\mu$ is high enough, the viscosity term in the above equation can prevent the system from developing shell-crossing singularities. Setting $\mu$ as constant, the above equation is Burgers' equation, which solutions are known analytically \cite{kevorkian}. Using the Cole-Hopf transform
\begin{equation}
    \tilde u=-2\mu\pdv{q_1}\ln\phi,
\end{equation}
Burgers' equation (with periodic boundary conditions) turns into a heat equation for $\phi$,
\begin{equation}
    \pdv{\phi}{b}=\mu\pdv[2]{\phi}{q_1}.
\end{equation}
This equation can be solved using Fourier analysis, and then inverting the Cole-Hopf transform leads to the general solution of (\ref{burgers}) with constant viscosity and initial condition $\tilde u(0,q_1)=\tilde U(q_1)$:
\begin{eqnarray}
    &\tilde u(q_1,b)
    =-2\mu\pdv{q_1}\ln \mathfrak{A}\nonumber\\
    &\mathfrak{A}=\dfrac{1}{\sqrt{4\pi\mu b}}
    \int_{-\infty}^\infty\exp \mathfrak{B}\dd s ;\nonumber\\
    &\mathfrak{B} = [-\dfrac{(q_1-s)^2}{4\mu b}-\dfrac12\int_0^s \tilde U(\xi)\dd\xi].
\end{eqnarray}
In order to study the constant $\mu$ approximation (which was the assumption in the ``adhesion approximation''), we will solve numerically Equation (\ref{burgers}), however, taking into account the time-dependence of $\mu$. The spatial dependence (coming from $\delta$) will not be considered for the time being. In this approximation, the viscosity coefficient $\mu(t)$ reads
\begin{equation}
    \mu(t)=3\beta\rho_{H,0}^2\dfrac{b}{a^4\dot b^2}.
\end{equation}
To compute this quantity, we need to provide the evolution of parameters $a$ and $b$ along with their derivatives. In a flat universe (with no curvature) and with the constraint that $a(t_{\rm i})=b(t_{\rm i})=1$ where $t_{\rm i}$ is the initial time, these read \cite{bildhaueretal}
\begin{subequations}
\begin{gather}
    a(\tau)=r^{1/3}\sinh^{2/3}(\omega\tau),\\
    h(\tau)=\dfrac{1}{a}\dv{a}{\tau}=\dfrac{2\omega}{3}\dfrac{1}{\tanh(\omega\tau)},\\
    b(\tau)=\dfrac{5}{6}\dfrac{r^{1/3}}{\sqrt{x}}B_x\qty(\dfrac{5}{6},\dfrac{2}{3}),\\
    \dv{b}{\tau}\,\!(\tau)=\dfrac{5}{2}r^{4/3}\dfrac{h}{a^3}\sqrt{x}\qty[x^{5/6}(1-x)^{-1/3}-\dfrac{1}{2}B_x\qty(\dfrac{5}{6},\dfrac{2}{3})],
\end{gather}
\end{subequations}
where $r=8\pi G\rho_{H,\rm i}/\Lambda$ is the matter to dark energy ratio in the Universe at initial time, $\tau=H_{\rm i} t$, and where $H_{\rm i}$ is the initial value of the Hubble parameter, $\omega=\sqrt{3\Lambda}/2H_{\rm i}$, and $x=1/(1+ra^{-3})$. $B_x(a,b)$ denotes the incomplete beta function,
\begin{equation}
    B_x(a,b)=\int_0^x t^{a-1}(1-t)^{b-1}\dd t.
\end{equation}

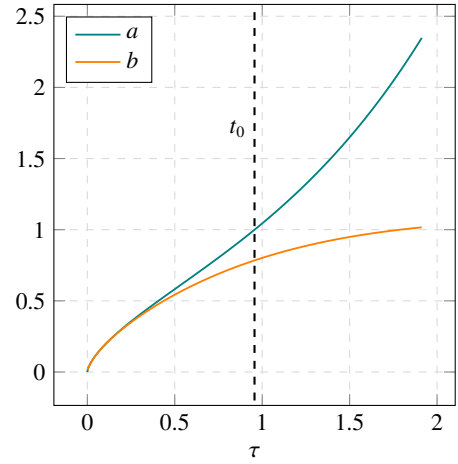
\begin{figure}[ht]
\centering
	\begin{tikzpicture}
		\begin{axis} [
			width = .8\linewidth,
			height = .8\linewidth,
			xlabel = $\tau$,
			grid = major,
			grid style = {dashed, gray!30},
			cycle list name=exotic,
			legend pos=north west,
			]
			\addplot+[mark=none, line width=.7pt] table[x=x, y=y1, col sep=semicolon] {ab_params.csv};
			\addplot+[mark=none, line width=.7pt] table[x=x, y=y2, col sep=semicolon] {ab_params.csv};
			\draw[line width=.8pt, dashed] (axis cs:0.956,-.2) -- (axis cs:0.956,2.65) node[pos=.67, left] {$t_0$}; 
			\legend{$a$, $b$}
		\end{axis}
	\end{tikzpicture}
\caption[Evolution of the parameters $a$ and $b$ from $0$ to $2t_0$.]{Evolution of the parameters $a$ and $b$ from $0$ to $2t_0$. Values of cosmological constants are taken from \cite{planck_collaboration}.} 
\label{fig:ab_params_1}
\end{figure}
\begin{figure}[ht]
\centering
	\begin{tikzpicture}
		\begin{axis} [
			width = .8\linewidth,
			height = .8\linewidth,
			xlabel = $\tau$,
			grid = major,
			grid style = {dashed, gray!30},
			cycle list name=exotic,
			ymax=5,
			]
			\addplot+[mark=none, line width=.7pt] table[x=x, y=y3, col sep=semicolon] {ab_params.csv};
			\addplot+[mark=none, line width=.7pt] table[x=x, y=y4, col sep=semicolon] {ab_params.csv};
			\draw[line width=.8pt, dashed] (axis cs:0.956,-.3) -- (axis cs:0.956,5) node[pos=.7, left] {$t_0$}; 
			\legend{$\dot a/a$, $\dot b/b$}
		\end{axis}
	\end{tikzpicture}
\caption[Evolution of the parameters $h=\dot a/a$ and $\dot b/b$ from $0$ to $2t_0$]{Evolution of the parameters $h=\dot a/a$ and $\dot b/b$ from $0$ to $2t_0$. Values of cosmological constants are taken from \cite{planck_collaboration}.} 
\label{fig:ab_params_2}
\end{figure}
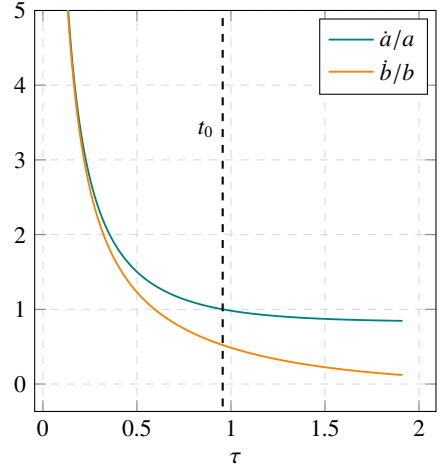
From now on, we will denote with an overdot the derivative with respect to $\tau$. With values of cosmological parameters at present time taken from \cite{planck_collaboration}, we derive their values at the initial time, chosen at the time of recombination, 
\begin{eqnarray}
    &H_{\rm i}=1.357 \times 10^6\ {\rm km}\;{\rm s}^{-1} \;{\rm Mpc}^{-1};\nonumber\\ &\Omega_{\mathrm{i},\Lambda}={2.49}\times 10^{-9}\ \  ; \ \ 
    \Omega_{\mathrm{i},m}=1-\Omega_{0,\Lambda};
\end{eqnarray}
the obtained evolution of $a,b$ and their derivatives is represented in Figures \ref{fig:ab_params_1} and \ref{fig:ab_params_2}, for $t\in[0,2t_0]$. As expected, we see that $b\sim a\sim t^{2/3}$ (as in an Einstein-de Sitter universe model) for $a\ll 1$, and that $b$ converges to a finite value when $a\gg 1$, whereas $a$ grows exponentially due to the dark energy term.
\subsection{Numerical simulation results}
In order to compare the different approximations that were made above with the exact solution of the nonperturbative equation (\ref{p1_beta}), we numerically integrate it and compare the predicted evolution at different times with the various approximations. All simulations start from the time of recombination, where the scale factor is taken to be unity. Every equation is integrated using implicit second-order finite-differences schemes \cite{kevorkian}. We first integrate both the exact and the linearized versions of Equation (\ref{p1_beta}) in Lagrangian coordinates (recall that the overdot is merely a partial time-derivative). The resulting $P_1(X_1,t)$ is then mapped to Eulerian Hubble-comoving space as $P_1(q_1,t)=P_1(X_1+P_1(X_1, t),t)$ for later comparison with the other simulations. Initial conditions are $P_1(X_1,t_{\rm i})=0$ in order to match the requirement $\vb X(t_{\rm i})=\vb x(t_{\rm i})$, and $\dot P_1(X_1,t_{\rm i})=U(X_1)$, where $U$ is the initial peculiar-velocity field. Boundary conditions are assumed periodic.\smallbreak
Then, Burgers' equation with time-dependent viscosity (\ref{burgers}) is integrated in Eulerian Hubble-comoving coordinates, where the total derivative $d_b$ reads $d_b=\partial_b+\tilde u/a\,\partial_{q_1}$. Initial conditions are $\tilde u(q_1,t_{\rm i})=\tilde U(q_1)$ and boundary conditions are assumed periodic. Finally, the same procedure is applied to solve Burgers' equation with constant viscosity, where the value of $\mu$ is taken as the average value of the time-dependent viscosity coefficient over the integration interval. However, there is always a degree of freedom in the definition of $\mu$, as the constant $\beta$ is \textit{a priori} unknown. In our simulations, the value of $\beta$ will be manually adjusted so that Burgers' solution coincides best with the exact solution. \smallbreak
The integration time interval is the same for all the simulations. Its end is determined by the time when $\delta$ locally reaches a magnitude that renders the nonlinear term in (\ref{p1_beta}) divergent. In order to realize shell-crossing for some coherence scale, the initial peculiar-velocity profile is taken of the simple form
\begin{equation}
    \tilde U(\hat q)=\sin\qty(\dfrac{2\pi\hat q}{L}),
\end{equation}
where $\tilde U,\hat q$ and $L$ are the dimensionless initial peculiar-velocity, spatial coordinate and length of the spatial domain of the simulation (see Appendix \ref{appB} for the definition of the dimensionless variables).
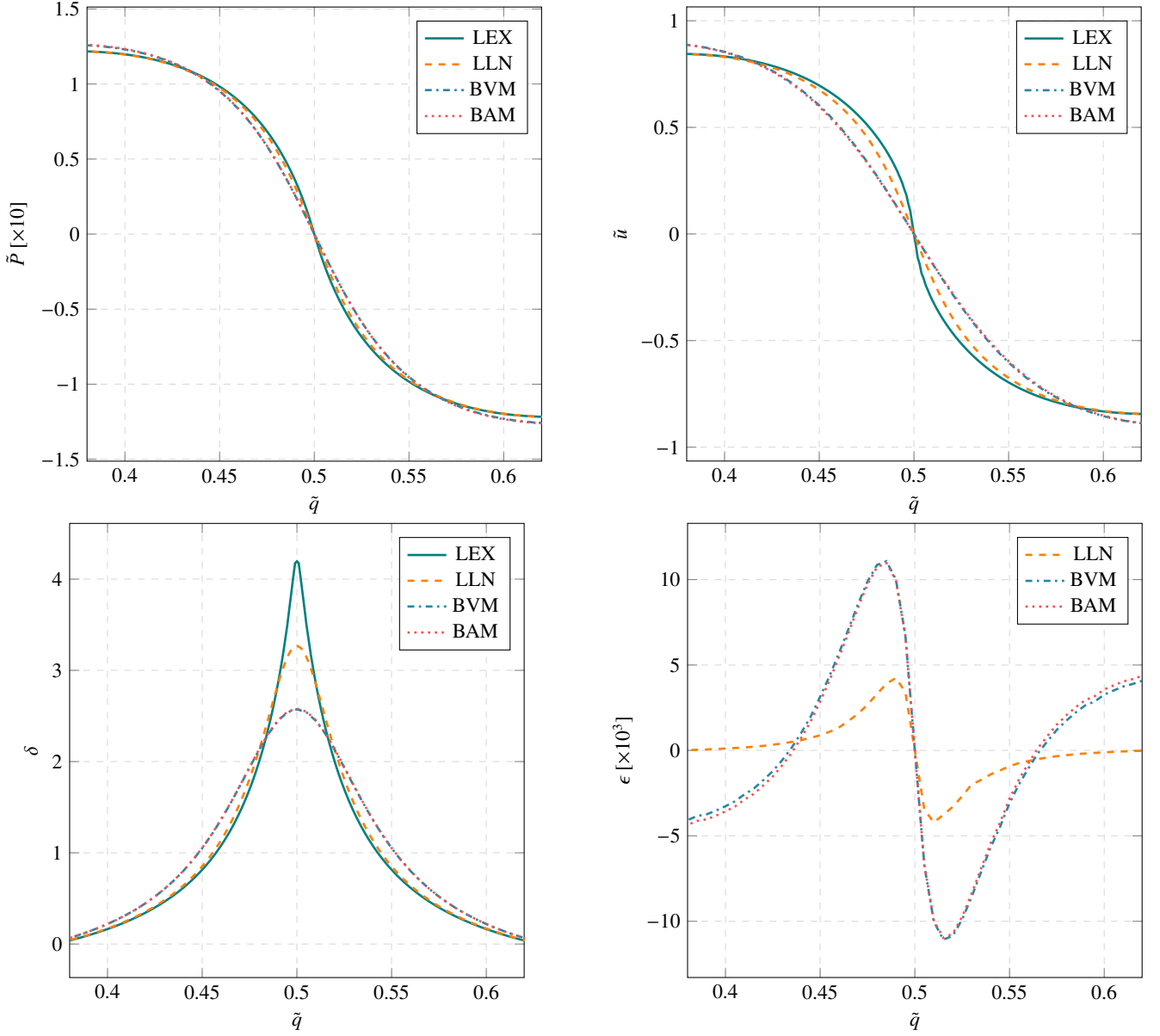
\begin{figure*}
    \centering
    \begin{subfigure}[t]{.48\linewidth}
    \centering
    	\begin{tikzpicture}
    		\begin{axis} [
    			width = \linewidth,
    			height = \linewidth,
    			xlabel = $\tilde q$,
     			ylabel = {$\tilde P\ [\times 10]$},
     			xmin = 0.38, xmax = 0.62,
    			grid = major,
    			grid style = {dashed, gray!30},
    			cycle list name=exotic,
    			legend pos=north east,
    			]
    			\addplot+[mark=none, line width=1pt] table[x=q_ex, y=P_ex, col sep=semicolon] {P_L=1.csv};
    			\addplot+[mark=none, line width=1pt, dashed] table[x=q_lin, y=P_lin, col sep=semicolon] {P_L=1.csv};
    			\addplot+[mark=none, line width=1pt, dashdotted] table[x=X, y=PBv, col sep=semicolon] {P_L=1.csv};
    			\addplot+[mark=none, line width=1pt, dotted] table[x=X, y=PBa, col sep=semicolon] {P_L=1.csv};
    			\legend{LEX, LLN, BVM, BAM}
    		\end{axis}
    	\end{tikzpicture}
    \label{fig:P_L=1}
    \end{subfigure}\hfill
    \begin{subfigure}[t]{.48\linewidth}
    \centering
	\begin{tikzpicture}
		\begin{axis} [
			width = \linewidth,
			height = \linewidth,
			xlabel = $\tilde q$,
 			ylabel = {$\tilde u$},
 			xmin = 0.38, xmax = 0.62,
			grid = major,
			grid style = {dashed, gray!30},
			cycle list name=exotic,
			legend pos=north east,
			]
			\addplot+[mark=none, line width=1pt] table[x=q_ex, y=u_ex, col sep=semicolon] {u_L=1.csv};
			\addplot+[mark=none, line width=1pt, dashed] table[x=q_lin, y=u_lin, col sep=semicolon] {u_L=1.csv};
			\addplot+[mark=none, line width=1pt, dashdotted] table[x=X, y=uBv, col sep=semicolon] {u_L=1.csv};
			\addplot+[mark=none, line width=1pt, dotted] table[x=X, y=uBa, col sep=semicolon] {u_L=1.csv};
			\legend{LEX, LLN, BVM, BAM}
		\end{axis}
	\end{tikzpicture}
    \label{fig:u_L=1}
    \end{subfigure}
    \begin{subfigure}[t]{.48\linewidth}
    \centering
    	\begin{tikzpicture}
    		\begin{axis} [
    			width = \linewidth,
    			height = \linewidth,
    			xlabel = $\tilde q$,
     			ylabel = {$\delta$},
     			xmin = 0.38, xmax = 0.62,
    			grid = major,
    			grid style = {dashed, gray!30},
    			cycle list name=exotic,
    			legend pos=north east,
    			]
    			\addplot+[mark=none, line width=1pt] table[x=q_ex, y=d_ex, col sep=semicolon] {delta_L=1.csv};
    			\addplot+[mark=none, line width=1pt, dashed] table[x=q_lin, y=d_lin, col sep=semicolon] {delta_L=1.csv};
    			\addplot+[mark=none, line width=1pt, dashdotted] table[x=X, y=dBv, col sep=semicolon] {delta_L=1.csv};
    			\addplot+[mark=none, line width=1pt, dotted] table[x=X, y=dBa, col sep=semicolon] {delta_L=1.csv};
			\legend{LEX, LLN, BVM, BAM}
    		\end{axis}
    	\end{tikzpicture}
    \label{fig:d_L=1}
    \end{subfigure}\hfill
    \begin{subfigure}[t]{.48\linewidth}
    \centering
		\begin{tikzpicture}
    		\begin{axis} [
    			width = \linewidth,
    			height = \linewidth,
    			xlabel = $\tilde q$,
     			ylabel = {$\epsilon\ [\times 10^3]$},
     			xmin = 0.38, xmax = 0.62,
    			grid = major,
    			grid style = {dashed, gray!30},
    			cycle list name=exotic,
    			legend pos=north east,
    			]
    			\pgfplotsset{cycle list shift=1}
    			\addplot+[mark=none, line width=1pt, dashed] table[x=X, y=lin, col sep=semicolon] {error_L=1.csv};
    			\addplot+[mark=none, line width=1pt, dashdotted] table[x=X, y=Bv, col sep=semicolon] {error_L=1.csv};
    			\addplot+[mark=none, line width=1pt, dotted] table[x=X, y=Ba, col sep=semicolon] {error_L=1.csv};
			\legend{LLN, BVM, BAM}
    		\end{axis}
	    \end{tikzpicture}
    \label{fig:e_L=1}
    \end{subfigure}
    \caption[Comparison of the three different approximations with the exact solution for the plane-symmetric collapse.]{Comparison of the three different approximations with the exact solution of Equation (\ref{p1_beta}). We adjusted $\tilde\beta={1.5}\times 10^{-2}$ such that all approximations agree best with the exact integration. Top left: rescaled perturbation $\tilde P$. Top right: rescaled peculiar-velocity $\tilde u$. Bottom left: density contrast $\delta$ (exact). Bottom right: error $\tilde P_{\rm exact}-\tilde P_{\rm approx}$. \textit{LEX = Lagrange exact, LLN = Lagrange linear, BVM = Burgers' equation with time-dependent viscosity, BAM = Burgers' equation with average viscosity.}}
    \label{fig:models_comp}
\end{figure*}
\subsubsection{Comparison of the different approximations}
First, all the equations are integrated over a spatial domain of length $L=1$, from a scale factor $a-1=0$ to a scale factor $a-1={1.01}\times 10^{-3}$. The value of $\tilde\beta$ (see Appendix \ref{appB}) was adjusted to $\tilde\beta={1.5}\times 10^{-2}$ such that all approximations agree best with the exact integration. Figure \ref{fig:models_comp} shows the results obtained for the perturbation field, the peculiar-velocity, and the density contrast right before shell-crossing happens (note the almost vertical tangent of the exact peculiar-velocity field), around $\tilde q={0.5}$ where the deviations between both models are the largest.\smallbreak
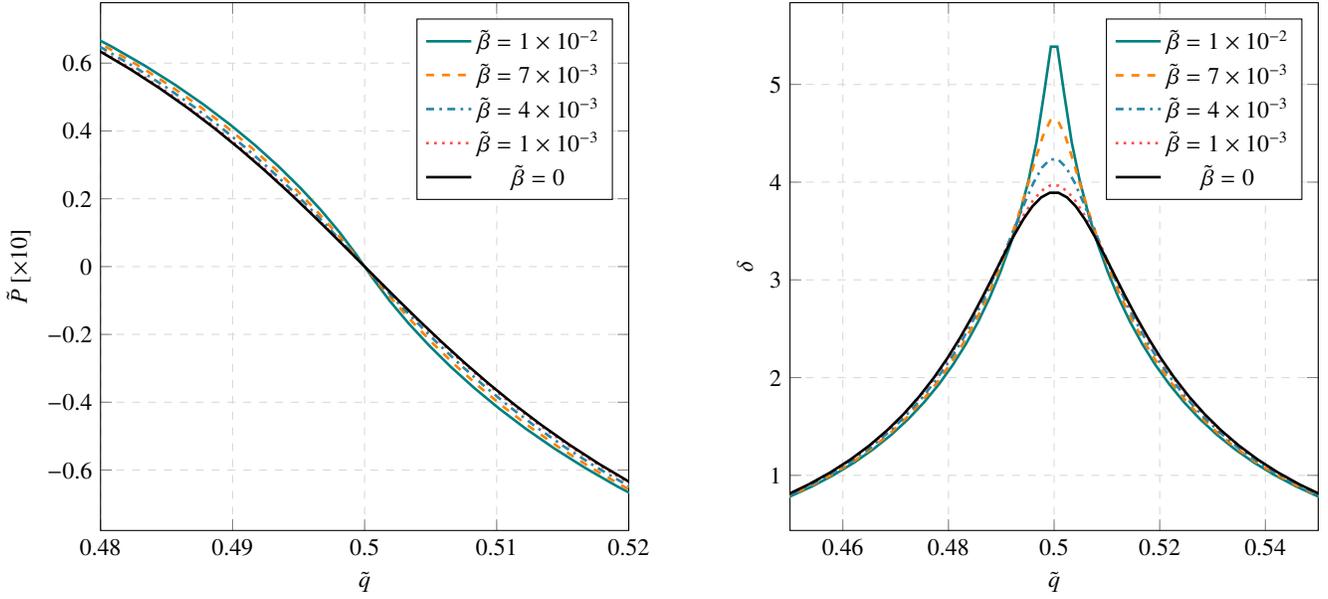
\begin{figure*}[ht]
    \centering
    \begin{subfigure}[t]{.48\linewidth}
    \centering
    	\begin{tikzpicture}
    		\begin{axis} [
    			width = \linewidth,
    			height = \linewidth,
    			xlabel = $\tilde q$,
     			ylabel = {$\tilde P\ [\times 10]$},
     			xmin = 0.48, xmax = 0.52,
    			grid = major,
    			grid style = {dashed, gray!30},
    			cycle list name=exotic,
    			legend pos=north east,
    			]
    			\addplot+[mark=none, line width=1pt] table[x=x, y=y1, col sep=semicolon] {all_bt=0.01.csv};
    			\addplot+[mark=none, line width=1pt, dashed] table[x=x, y=y1, col sep=semicolon] {all_bt=0.007.csv};
    			\addplot+[mark=none, line width=1pt, dashdotted] table[x=x, y=y1, col sep=semicolon] {all_bt=0.004.csv};
    			\addplot+[mark=none, line width=1pt, dotted] table[x=x, y=y1, col sep=semicolon] {all_bt=0.001.csv};
    			\addplot+[mark=none, line width=1pt, color=black] table[x=x, y=y1, col sep=semicolon] {all_bt=0.csv};
    		\legend{$\tilde\beta=\num{1e-2}$, $\tilde\beta=\num{7e-3}$, $\tilde\beta=\num{4e-3}$, $\tilde\beta=\num{1e-3}$, $\tilde\beta=0$}
    		\end{axis}
    	\end{tikzpicture}
    \end{subfigure}\hfill
    \begin{subfigure}[t]{.48\linewidth}
    \centering
    	\begin{tikzpicture}
    		\begin{axis} [
    			width = \linewidth,
    			height = \linewidth,
    			xlabel = $\tilde q$,
     			ylabel = {$\delta$},
     			xmin = 0.45, xmax = 0.55,
    			grid = major,
    			grid style = {dashed, gray!30},
    			cycle list name=exotic,
    			legend pos=north east,
    			]
			\addplot+[mark=none, line width=1pt] table[x=x, y=y3, col sep=semicolon] {all_bt=0.01.csv};
			\addplot+[mark=none, line width=1pt, dashed] table[x=x, y=y3, col sep=semicolon] {all_bt=0.007.csv};
			\addplot+[mark=none, line width=1pt, dashdotted] table[x=x, y=y3, col sep=semicolon] {all_bt=0.004.csv};
			\addplot+[mark=none, line width=1pt, dotted] table[x=x, y=y3, col sep=semicolon] {all_bt=0.001.csv};
			\addplot+[mark=none, line width=1pt, color=black] table[x=x, y=y3, col sep=semicolon] {all_bt=0.csv};
    		\legend{$\tilde\beta=\num{1e-2}$, $\tilde\beta=\num{7e-3}$, $\tilde\beta=\num{4e-3}$, $\tilde\beta=\num{1e-3}$, $\tilde\beta=0$}
    		\end{axis}
    	\end{tikzpicture}    
    \end{subfigure}
    \caption[Perturbation and density contrast fields obtained for various values of $\tilde\beta$.]{Perturbation and density contrast fields obtained for various values of the dimensionless parameter $\tilde\beta$ at a scale factor $a-1={1.4}\times 10^{-1}$. Plots with $\tilde\beta=0$ are added for reference and correspond to the dust model.}
    \label{fig:size_comp}
\end{figure*}
A first observation to make is that the results obtained within the ``adhesion approximation'' and Burgers' equation are almost identical either with variable or average viscosity. However, no conclusion can be drawn here, as the time interval of the simulation is so small that the viscosity parameter $\mu(t)$ almost does not change at all during that interval. The bottom-right panel of Figure \ref{fig:models_comp} shows the local error $\epsilon=P_{\rm exact}-P_{\rm approx}$ between each approximation of the perturbation field and its exact solution. Overall, all three approximations are accurate within a local error under 1\% almost everywhere. However, the Lagrange-linear approximation astonishingly exhibits a smaller error than that of the ``adhesion approximation'', which may show that the restricting assumptions underlying the latter are stronger than those of the ``mere'' Lagrange-linearization of the exact equation (\ref{p1_beta}). To reduce the number of approximations made in this case, one may try to use the divergence equation (\ref{ejn_pecd}) with the slaving condition (\ref{slaving}) to write $\delta$ as a function of $u$, leading to the conservation law
\begin{equation}\label{burgers+}
    \pdv{\tilde u}{b}=\pdv{q}\qty[\mu(t)\pdv{\tilde u}{q}-\dfrac{1}{2}\tilde u^2-\dfrac{\mu(t)b}{2}\qty(\pdv{\tilde u}{q})^2],
\end{equation}
instead of Burgers' equation. The added nonlinearity might help improve the description of shell-crossing, which is where all the approximations appear to break down, as shown by the density contrast plots. \smallbreak
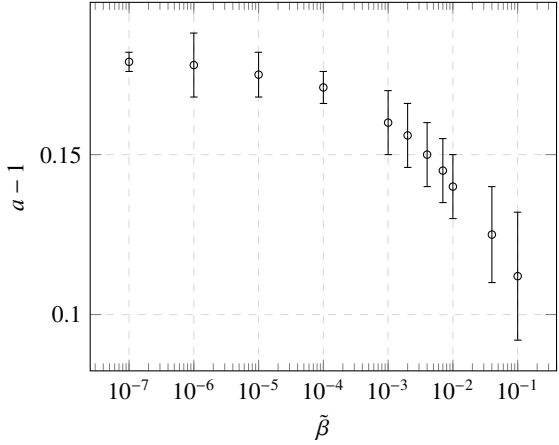
\begin{figure}[ht]
    \centering
    \begin{tikzpicture}
        \begin{semilogxaxis} [
			width = .9\linewidth,
			height = .75\linewidth,
			ylabel = {$a-1$},
			xlabel = {$\tilde\beta$},
			grid = major,
			grid style = {dashed, gray!30},
            error bars/y dir=both,
            error bars/y explicit,
			]
			\addplot[mark=o,
                mark size=1.4pt,
                only marks,] table[x=bt, y=a, y error=ua, col sep=semicolon] {sc_time.csv};
		\end{semilogxaxis}
	\end{tikzpicture}
    \caption{Scale factor at shell-crossing time for various values of the $\tilde\beta$ dimensionless parameter.}
    \label{fig:tsc}
\end{figure}
\subsubsection{Impact of the cloud's initial parameters}
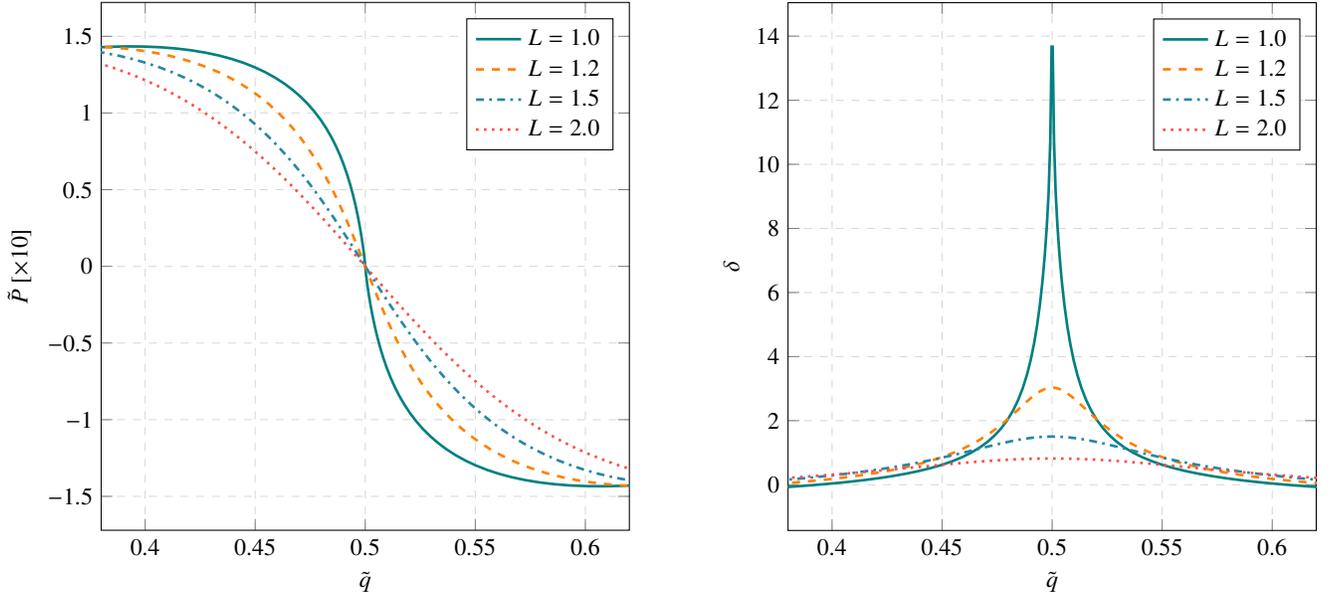
\begin{figure*}[ht]
    \centering
    \begin{subfigure}[t]{.48\linewidth}
    \centering
    	\begin{tikzpicture}
    		\begin{axis} [
    			width = \linewidth,
    			height = \linewidth,
    			xlabel = $\tilde q$,
     			ylabel = {$\tilde P\ [\times 10]$},
     			xmin = 0.38, xmax = 0.62,
    			grid = major,
    			grid style = {dashed, gray!30},
    			cycle list name=exotic,
    			legend pos=north east,
    			]
    			\addplot+[mark=none, line width=1pt] table[x=x, y=y1, col sep=semicolon] {all_L=1.csv};
    			\addplot+[mark=none, line width=1pt, dashed] table[x=x, y=y1, col sep=semicolon] {all_L=1.2.csv};
    			\addplot+[mark=none, line width=1pt, dashdotted] table[x=x, y=y1, col sep=semicolon] {all_L=1.5.csv};
    			\addplot+[mark=none, line width=1pt, dotted] table[x=x, y=y1, col sep=semicolon] {all_L=2.csv};
    			\legend{$L=1.0$, $L=1.2$, $L=1.5$, $L=2.0$}
    		\end{axis}
    	\end{tikzpicture}
    \label{fig:dP_L=1}
    \end{subfigure}\hfill
    \begin{subfigure}[t]{.48\linewidth}
    \centering
    	\begin{tikzpicture}
    		\begin{axis} [
    			width = \linewidth,
    			height = \linewidth,
    			xlabel = $\tilde q$,
     			ylabel = {$\delta$},
     			xmin = 0.38, xmax = 0.62,
    			grid = major,
    			grid style = {dashed, gray!30},
    			cycle list name=exotic,
    			legend pos=north east,
    			]
			\addplot+[mark=none, line width=1pt] table[x=x, y=y3, col sep=semicolon] {all_L=1.csv};
			\addplot+[mark=none, line width=1pt, dashed] table[x=x, y=y3, col sep=semicolon] {all_L=1.2.csv};
			\addplot+[mark=none, line width=1pt, dashdotted] table[x=x, y=y3, col sep=semicolon] {all_L=1.5.csv};
			\addplot+[mark=none, line width=1pt, dotted] table[x=x, y=y3, col sep=semicolon] {all_L=2.csv};
    		\legend{$L=1.0$, $L=1.2$, $L=1.5$, $L=2.0$}
    		\end{axis}
    	\end{tikzpicture}
    \label{fig:dd_L=1}
    \end{subfigure}
    \caption{Perturbation and density contrast fields obtained for various initial domain sizes $L$ with the parameter $\tilde\beta=10^{-3}$ at a scale factor $a-1={1.6}\times 10^{-1}$.}
    \label{fig:L_comp}
\end{figure*}

The value of the dimensionless parameter $\tilde\beta$ entering in all the collapse equations plays a significant role for the timescale of the collapse. Indeed, $\tilde\beta$ is linked to the cloud's initial Jeans' length and density via
\begin{equation}
	\tilde{\beta}=\frac{3\beta\rho_{H,\rm i}}{4\pi GL_{J,\rm i}^2},
\end{equation}
which means that large values of $\tilde\beta$ correspond to denser clouds, which should therefore collapse faster than larger and more dilute ones. Figure \ref{fig:size_comp} shows the perturbation and density contrast fields obtained with the Lagrange exact model, integrated from $a-1=0$ to $a-1={1.4}\times 10^{-1}$, with the parameter $\tilde\beta$ ranging from \num{1e-3} to \num{1e-2}. It appears clearly that the smaller the value of $\tilde\beta$, the later shell-crossing happens, with the maximum density contrast decreasing with decreasing $\tilde\beta$.\smallbreak
In order to study this dependence in more detail, we integrate the Lagrange exact model until shell-crossing with varying values of $\tilde\beta$ within $10^{-7}$ and $10^{-1}$. 
Figure \ref{fig:tsc} shows the scale factor at which shell-crossing happens for each value of $\tilde\beta$ on a semi-logarithmic scale. For very low values of $\tilde\beta$, the effect of velocity dispersion is completely negligible and the cloud collapses as in the dust model, which results in a constant value of $a$ at shell-crossing. The asymptotic value of $a$ is in agreement with the value $a-1=\num{1.792(2)e-1}$ obtained when taking $\tilde\beta=0$. Higher values of $\tilde\beta$ lead, however, to an earlier shell-crossing, which confirms the qualitative observations made in Figure \ref{fig:size_comp}. Our numerical integration code did not allow us to explore values of $\tilde\beta$ greater that $10^{-1}$.\smallbreak
Another important parameter is the initial size, $L$, of the cloud. Indeed, smaller clouds should collapse faster than larger ones. Figure \ref{fig:L_comp} shows the perturbation and density contrast fields obtained with the Lagrange exact model, integrated from $a-1 = 0$ to $a-1 =\num{1.6e-1}$, with the parameter $\tilde\beta=10^{-3}$, for various initial cloud sizes. In the plots, the domain size has been rescaled to fit within $q\in[0,1]$ so that the results can be compared. Clearly, shell-crossing happens later the larger the cloud. In order
to quantify this dependence, we integrate the Lagrange exact
model until shell-crossing with various initial domain sizes $L\in[1,100]$ for different values of $\tilde\beta$. Figure \ref{fig:tscL} shows the scale factor at which shell-crossing happens for
each value of $L$ on a logarithmic scale. It appears that, for both small and large values of $L$, the dependence is of power-law type, with the impact of $\tilde\beta$ being mostly relevant for small clouds. A power law of the form $a-1 = a_0L^r$ is then fitted to the obtained data for both $L\in[1,5]$ and $L\in[40,100]$, for each value of $\tilde\beta$. Results are gathered in Table \ref{tab:coefs}. All values of $\tilde\beta$ lead to similar $L$-dependence, with an exponent of $r\approx\num{1.2}$ for small clouds and $r\approx1$ for large clouds. For the latter, the behavior becomes independent of the value of $\tilde\beta$. This indicates that the effect of velocity dispersion is mostly significant in small matter clouds, and tends to be negligible in larger ones. In small clouds, the effect of velocity dispersion can be seen in the prefactor $a_0$, which decreases rapidly with increasing $\tilde\beta$. However, no clear analytical dependence between $a_0$ and $\tilde\beta$ can be inferred from the available data. Also, our current numerical simulation does not allow us to gather data for very large clouds $L\gtrsim100$.  
\begin{figure}[ht]
    \centering
    \begin{tikzpicture}
        \begin{loglogaxis} [
			width = .9\linewidth,
			height = .75\linewidth,
			ylabel = {$a-1$},
			xlabel = {$L$},
			grid = minor,
			grid style = {dashed, gray!30},
            error bars/y dir=both,
            error bars/y explicit,
            cycle list name=exotic,
            legend pos=south east,
			]
			\addplot+[mark=o,
                mark size=1.4pt,
                only marks,] table[x=L, y=a0, y error=u0, col sep=semicolon] {sc_time_L.csv};
            \addplot+[mark=o,
                mark size=1.4pt,
                only marks,] table[x=L, y=a2, y error=u2, col sep=semicolon] {sc_time_L.csv};
			\pgfplotsset{cycle list shift=1}
            \addplot+[mark=o,
                mark size=1.4pt,
                only marks,] table[x=L, y=a1, y error=u1, col sep=semicolon] {sc_time_L.csv};
                \addplot[samples=2, domain=1:7, dashed] {0.172*x^1.24};
                \addplot[samples=2, domain=30:100, dashed] {0.3*x^1.04};
                \addplot[samples=2, domain=1:7, dashed] {0.13*x^1.26};
                \addplot[samples=2, domain=1:7, dashed] {0.107*x^1.23};
                \legend{$\tilde\beta=0$,$\ \tilde\beta=10^{-2}$,$\ \tilde\beta=10^{-1}$}
		\end{loglogaxis}
	\end{tikzpicture}
    \caption{Scale factor at shell-crossing time for various initial cloud sizes $L$. Dashed lines correspond to the fitted power laws for choices of $\tilde\beta$ whose parameters are given in table \ref{tab:coefs}.}
    \label{fig:tscL}
\end{figure}
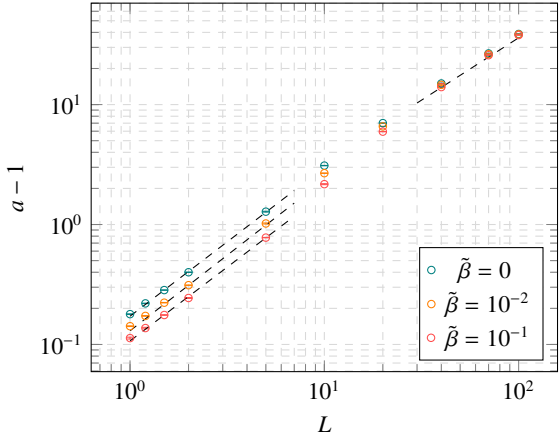
\begin{table}[ht]
\begin{tabular}{c|cccc}
\multirow{2}{*}{$\tilde\beta$} & \multicolumn{2}{c}{$L\in[1,5]$} & \multicolumn{2}{c}{$L\in[40,100]$} \\
                               & $a_0$           & $r$           & $a_0$            & $r$             \\ \hline \\
$0$                            & \num{0.172(9)}  & \num{1.24(4)} & \num{0.3(1)}     & \num{1.04(7)}   \\
$10^{-2}$                      & \num{0.13(1)}   & \num{1.26(5)} & \num{0.29(9)}    & \num{1.06(7)}   \\
$10^{-1}$                      & \num{0.107(8)}  & \num{1.23(5)} & \num{0.25(3)}    & \num{1.08(3)}  
\end{tabular}
\caption{Power law coefficients for the dependence of the shell-crossing time on the initial cloud size $L$.}
\label{tab:coefs}
\end{table}
%
\section{Three-dimensional gravitational collapse with velocity dispersion}
\label{3D}
In this section we aim at extending the results previously obtained in the plane-symmetric case to a more general case without any particular symmetry, in the spirit of \cite{doroshkevichetal}.
\subsection{Maximal anisotropy and estimate of importance of derivatives}
Three-dimensional collapsing fluids were first understood in the case of spherically-symmetric systems, which allowed for nonlinear analytical developments. However, the work of Zel'dovich (e.g., among many others, \cite{zeldovich70a, zeldovich70b}) has shown that these systems are not spherically symmetric, but instead \textit{pancake}-like, exhibiting maximal anisotropy in one particular direction. Another way of saying this is to consider the scalar invariants of the Lagrangian perturbation tensor $P_{i|j}$ (that have been defined previously). Maximal anisotropy implies that this tensor has one dominating eigenvalue, which we may call $\lambda_1$: $\lambda_1\gg\lambda_2,\lambda_3$, which implies that the first scalar invariant is dominating the other two: $\mathrm{I}\gg\mathrm{II},\mathrm{III}$, see the appendix of \cite{vigneron:darkmatter} for a recent collection of arguments and numerical tests.\smallbreak
This approximation only takes into account the qualitative kinematics of the collapse, without any consideration on the magnitude of the individual coefficients $P_{i|j}$. However, it can be shown in the context of general relativity that both the perturbation gradient $P_{i|j}$ and its first derivatives $P_{i|jk}$, respectively related to the perturbations of the 3-metric $g_{ij}$ and the spatial connection coefficients $\Gamma^i_{\,jk}$, can be considered small in regular systems (i.e., without singularities) \cite{buchert_geometrical_2009}. However, the second derivatives (related to the spatial curvature) are not typically small in those systems. Although in Newtonian cosmology spacetime is flat, the analogy and the two-scale argument of \cite{buchert_geometrical_2009} prevails algebraically in the Lagrangian representation of the Newtonian system with the Lagrangian metric, $g_{k\ell}=\delta_{ij} f^i_{\ |k}f^j_{\ | \ell}$, and the same arguments allow to estimate the smallness in magnitude of the different levels of derivatives. The reader may consult the explicit presentation of this correspondence in \cite{correspondence}.\smallbreak
Taking into account both of these approximations (i.e. the perturbation gradient itself and its first derivatives are small), one can simplify the expressions involved in Equations (\ref{lagr_p}) by neglecting all nonlinear terms in the perturbations or their first derivatives, along with terms proportional to II and III, but keeping second derivatives. 
\subsection{Field equations}
Using the hypotheses made above, the remaining terms involved in (\ref{lagr_p}) read:
\begin{subequations}
\begin{gather}
    J=a^3(1+\mathrm{I})=a^3(1+P_{k|k}),\\
    \J(J,f_i,f_j)=a^5\epsilon_{kij}\mathrm{I}_{|k},\\
    \J(\J(J,f_i,f_j),f_i,f_j)=2a^7\mathrm{I}_{|kk},\\
    \epsilon_{kij}\J(\ddot f_k,f_i,f_j)=6\dfrac{\ddot a}{a}a^3(1+\mathrm{I})+4Ha^3\dot{\mathrm{I}}+2a^3\ddot{\mathrm{I}}.
\end{gather}
Moreover, considering that the anisotropy $\Delta_{ij}$ is bound to the same level of approximation imposed on the whole perturbation gradient, the last term evaluates to
\begin{equation}
    \epsilon_{kij}\epsilon_{pqr}\J(\J(\Delta_{kp},f_q,f_r),f_i,f_j)=4a^4\Delta_{kp|pk}.
\end{equation}
\end{subequations}
Injecting all the above expressions into the field equation (\ref{lagr_pb}), one obtains:
\begin{eqnarray}
    &3\dfrac{\ddot a}{a}+2H\dfrac{\dot{\I}}{1+\I}+\dfrac{\ddot{\I}}{1+\I}-\dfrac{\alpha'}{a^2}\dfrac{\I_{|kk}}{(1+\I)^3}+\dfrac{a}{C_H}\dfrac{\Delta_{kp|pk}}{1+\I}\nonumber\\
    &=\Lambda-4\pi G\qty(\dfrac{C_H}{a^3(1+\I)}+3p_H).
\end{eqnarray}
Injecting Friedmann's equation (\ref{fried}) and linearizing with respect to $\I$ then yields:
\begin{equation}\label{3d_I}
    \ddot{\I}+2H\,\dot{\I}-4\pi G\dfrac{C_H}{a^3}\I=\dfrac{\alpha'}{a^2}(1-2\I)\laplacian_0\,\I+\dfrac{a}{C_H}\Delta_{kp|pk}.
\end{equation}
The form of this equation is reassuring, as the left-hand side is the well-known time-derivative operator appearing in every perturbative development discussed in Section \ref{par:motiv}. The Laplacian term on the right-hand side is also the same as in Lagrangian-linear perturbation theory, reminiscent of Equation (\ref{lagr_lin}). The remaining term encodes the effects of the anisotropy of the velocity dispersion. It is to be noted though, that the Lagrangian Laplacian appearing in the right-hand side is only Lagrangian due to the linearization in the first invariant $\rm I$ and its gradient. Quadratic terms in $\rm I$ and all Lagrange-nonlinear terms neglected above sum up to the full Eulerian Laplacian.\smallbreak
Now we proceed by determining equations of state for the isotropic pressure $p$ and the anisotropic dispersion $\Delta_{ij}$, in the form $p=\alpha(\rho)$ and $\Delta_{ij}=\beta_{ij}(\rho,\vb v)$. The (full) velocity field $\vb v$, being the source of anisotropy in the system, the anisotropic part of the velocity dispersion tensor must be a function of the latter, and not only of the density. At this point it is useful to introduce the kinematic decomposition of the full Eulerian velocity gradient $v_{i,j}$:
\begin{equation}
    v_{i,j}=\dfrac{1}{3}\theta\delta_{ij}+\sigma_{ij}+\omega_{ij},
\end{equation}
where $\theta=v_{k,k}=-\dot\rho/\rho$ is the rate of expansion, $\sigma_{ij}$ the shear tensor (symmetric and traceless) and $\omega_{ij}$ the vorticity tensor (antisymmetric). Using the definition (\ref{peculiara}) of the peculiar-velocity, we can write this decomposition for the peculiar-velocity as
\begin{equation}
    \dfrac{1}{a}\pdv{u_i}{q_j}=\dfrac13(\theta-3H)\delta_{ij}+\sigma_{ij}+\omega_{ij}.
\end{equation}
Injecting this decomposition into the Euler-Jeans equation (\ref{ejn_pecc}), we obtain:
\begin{equation}
    \dv{t}\qty(p\delta_{ij}+\Delta_{ij})+\dfrac{5}{3}\theta\; \qty(p\delta_{ij}+\Delta_{ij})=-2\qty[p\sigma_{ij}+\sigma_{ik}\Delta_{kj}].
\end{equation}
Taking the trace of this equation and taking into account that both $\sigma_{ij}$ and $\Delta_{ij}$ are traceless yields
\begin{equation}
\label{EOS}
    3\dfrac{\dot p}{p}+5\theta=-2\sigma_{ik}\Delta_{ki}.
\end{equation}
This has once again the expected form, leading in the isotropic limit $\Delta_{ij}=0$ to the classical polytropic solution $p\propto\rho^{5/3}$ reminiscent of an adiabatic ideal gas, which is a correct analogy if we consider our system as a dilute fluid of massive particles, the difference being the kinetic pressure source which does not originate from collisions between particles but from (isotropic) velocity dispersion. However, here the anisotropic part of the velocity dispersion has a feedback effect on the isotropic pressure, which alters the equation of state and makes $p$ become also velocity-dependent (\textit{via} the shear tensor).\footnote{With Equation \eqref{EOS} we also recover the equation of state of the background dispersion for vanishing shear \cite{ehlersrienstra}, \cite{buchert:adhesive}.} \smallbreak
Here, in accordance with our approximation that the perturbation gradient $P_{i|j}$ is small, we make the approximation that the peculiar-velocities are small, which in turn implies that also both the shear and the anisotropic velocity dispersion are small. Keeping only order 1 terms in both $\sigma_{ij}$ and $\Delta_{ij}$ in the previous equations leads to
\begin{subequations}
\begin{gather}
    \dv{t}\qty(p\delta_{ij}+\Delta_{ij})+\dfrac{5}{3}\theta\;\qty(p\delta_{ij}+\Delta_{ij})=-2p\sigma_{ij},\\
    \dfrac{\dot p}{p}+\dfrac{5}{3}\theta=0,
\end{gather}
\end{subequations}
and the trace equation is immediately solved by $p=\kappa\rho^{5/3}$, where $\kappa$ can be set constant with appropriate initial conditions. Injecting back this solution into the first equation then yields
\begin{equation}
    \dot\Delta_{ij}+\dfrac{5}{3}\theta\Delta_{ij}=-2\kappa\rho^{5/3}\sigma_{ij}.
\end{equation}
It is interesting here to see that $\Delta_{ij}$ indeed depends on the total velocity field $\vb v$, and not only on the peculiar-velocity, despite the Hubble-velocity being isotropic. For the sake of consistency, we translate this equation into the Lagrangian picture. Using the integrals (\ref{lagr_int}) for $\rho$ and $\vb v$, we obtain the following expressions:
\begin{subequations}
\begin{gather}
    \theta=\dfrac{\dot J}{J},\\
    p=\kappa\dfrac{C_H}{J^{5/3}},\\
    \sigma_{ij}=\dfrac{1}{4J}\qty(\epsilon_{jk\ell}\J(\dot f_i,f_k,f_\ell)+\epsilon_{ik\ell}\J(\dot f_j,f_k,f_\ell)).
\end{gather}
\end{subequations}
Within our approximations on the importance of different levels of derivatives, these reduce to
\begin{subequations}
\begin{gather}
    \theta=3H+\dot{\I},\\
    p=\dfrac{C_H^{5/3}}{a^5}\qty(1-\dfrac{5}{3}\I),\\
    \sigma_{ij}=\dfrac{1}{2}\qty(\dot P_{i|j}+\dot P_{j|i})-\dfrac13\dot{\I}\delta_{ij},
\end{gather}
\end{subequations}
and we are finally left with the following system:
\begin{subequations}\label{lagr_3d_approx}
\begin{gather}
    \ddot{\I}+2H\,\dot{\I}-4\pi G\dfrac{C_H}{a^3}\I\nonumber\\
    =\dfrac53\dfrac{\kappa C_H^{2/3}}{a^4}\qty(1-\dfrac{8}{3}\I)\laplacian_0\,\I+\dfrac{a}{C_H}\Delta_{kp|pk},
    \label{lagr_3d_approxa}\\
    \dot{\Delta}_{ij}+5H\Delta_{ij}=-\dfrac{\kappa C_H^{5/3}}{a^5}\qty(\dot P_{i|j}+\dot P_{j|i}-\dfrac{2}{3}\dot{\I}\delta_{ij}).
     \label{lagr_3d_approxb}
\end{gather}
Yet this system is not closed, as it contains nine unknowns (three coefficients of the perturbation vector $\vb P$ and six coefficients of the symmetric anisotropic velocity dispersion tensor) and only eight equations (one scalar equation (\ref{lagr_3d_approxa}), six equations (\ref{lagr_3d_approxb}) plus the traceless condition on $\Delta_{ij}$). In order to close the system, one may use the first set of Lagrangian equations (\ref{lagr_pa}), which gives three more equations in the form
\begin{equation}
    \grad_0\times\qty[\ddot{\vb P}+2H\dot{\vb P}+\dfrac{a}{C_H}\grad_0\cdot\vb\Delta]={\bf 0}.
\end{equation}
\end{subequations}
This set of equations can be viewed as evolution equations for the whole perturbation gradient tensor $P_{i|j}$, decomposed into its trace, traceless symmetric and antisymmetric parts. Using the notations of \cite{rza4} in the GR setting, we write the Lagrange-Newton counterpart,
\begin{equation}
    P_{i|j}=\dfrac{1}{3}P\delta_{ij}+\mathfrak{S}_{ij}+\mathfrak{P}_{ij},
\end{equation}
where we defined $P=P_{k|k}=\I$, $\mathfrak{S}_{ij}=P_{(i|j)}-P/3\,\delta_{ij}$ and $\mathfrak{P}_{ij}=P_{[i|j]}$. The previous equations then finally read:
\begin{subequations}\label{3d_Pij}
\begin{gather}
    \ddot P+2H\dot P-4\pi G\rho_H P\nonumber\\
    =\dfrac{5}{3}\dfrac{\kappa\rho_H^{2/3}}{a^2}\qty(1-\dfrac{8}{3}P)\laplacian_0P+\dfrac{1}{a^2\rho_H}\Delta_{kp|pk},
    \label{3d_Pija}\\
    \kappa\rho_H^{5/3}\dot{\mathfrak{S}}_{ij}+\dot\Delta_{ij}+5H\Delta_{ij}=0,\label{3d_Pijb}\\
    \epsilon_{ijk}\qty[\ddot{\mathfrak{P}}_{kj}+2H\dot{\mathfrak{P}}_{kj}+\dfrac{1}{a^2\rho_H}\Delta_{kp|pj}]=0.
    \label{3d_Pijc}    
\end{gather}
\end{subequations}
\subsection{Discussion}
Looking at the previous set of equations governing the evolution of the perturbation gradient, it is interesting to note the similarity with the plane-symmetric case, and in particular Equations (\ref{p1_delta}) and (\ref{p1_beta}). Indeed, linearizing the former with respect to $P_1,P_{1|1}$ and $P_{1|11}$ as prescribed by the GR analogy, we recover an equation much similar to (\ref{3d_Pija}). The main difference between the two comes from the equation of state, which for the plane-symmetric case was $p\propto\rho^3$ and in the approximate 3D case is $p\propto\rho^{5/3}$. This only changes the coefficient in front of the Laplacian, which we may call the \textit{speed of sound}. Indeed, for such a self-gravitating fluid, the speed of sound waves can be defined as 
\begin{equation}
    c_s^2=\dv{p}{\rho}=\alpha'(\rho)=\dfrac{5}{3}\kappa\rho_H^{2/3},
\end{equation}
which allows to interpret Equation (\ref{3d_Pij}) as a nonlinear wave equation, rewritten in the form
\begin{eqnarray}
    &\ddot P-\dfrac{c_s^2}{a^2}\laplacian_0 P+2H\dot P-4\pi G\rho_H P\nonumber\\
    &=\dfrac{8}{3}\dfrac{c_s^2}{a^2}P\laplacian_0P+\dfrac{1}{\rho_H a^2}\Delta_{kp|pk}\ .
\end{eqnarray}
The first two terms form the d'Alembertian of $P$ which drives the evolution of waves in the fluid. The other terms are a damping term in $\dot P$, source terms in $P$ and $\Delta$, and a nonlinear term in $P\laplacian_0P$. The linearized trace equation thus describes the propagation of ``sound waves'' in a self-gravitating fluid with velocity dispersion. However, the nonlinearity that appears here is a highly approximated result where higher-order nonlinear terms in $P$ and $\grad_0P$ were neglected. \smallbreak
The other main difference with the plane-symmetric case is that the off-diagonal terms of the perturbation gradient are \textit{a priori} nonzero. The expression for $\Delta_{ij}$ is then not immediate, and is instead given by the evolution equation (\ref{3d_Pijb}). The nonzero anisotropic dispersion then becomes a source term for the vorticity in Equation (\ref{3d_Pijc}). Indeed, one can show that the vorticity vector $\omega_i=1/2\,\epsilon_{ijk}\omega_{kj}$ (where $\omega_{ij}=v_{[i,j]}$ is the vorticity tensor as defined previously) can be expressed in the linearized Lagrangian picture as 
\begin{equation}
    \omega_i=\dfrac{1}{2}\epsilon_{ijk}\dot{\mathfrak{P}}_{kj},
\end{equation}
so that Equation (\ref{3d_Pijc}) reads:
\begin{equation}
    \dot\omega_i+2H\omega_i+\dfrac{2}{a^2\rho_H}\epsilon_{ijk}\Delta_{kp|pj}=0.
\end{equation}
(A perturbative expansion for the vorticity field's evolution equation is given in \cite{ruth:vorticity}.)
Now let us look at an interesting limiting case. Going back to our main hypothesis that the perturbation gradient is dominated by one eigenvalue, which can be time and space-dependent, we consider that \textit{locally}, the eigendirections of the perturbation gradient and the velocity dispersion tensor $\pi_{ij}$ are coincident, so that we can diagonalize both of them simultaneously. This way, we can locally restrict ourselves to a case where $\pi_{ij}$ is also dominated by one eigenvalue, which physically corresponds to the velocity dispersion being significant only in the direction of the collapse. This \textit{locally 1D} approximation leads to a situation that is similar to the plane-symmetric case, where we consider the velocity dispersion tensor to only have one significant term on its diagonal, with all other terms being negligible along with their derivatives. The vorticity equation then becomes the usual linearized Kelvin-Helmholtz transport equation \cite{serrin}, \cite{ehlersbuchert},
\begin{subequations}
\begin{equation}
    \dot{\boldsymbol{\omega}}+2H\boldsymbol{\omega}={\bf 0},
\end{equation}
which is solved by the linearized exact Cauchy integral
\begin{equation}
    \boldsymbol{\omega}=\dfrac{\vb\Omega}{a^2}\qc \vb\Omega=\boldsymbol{\omega}(\vb X,t=t_{\rm i}).
\end{equation}
\end{subequations}
\section{Conclusion and outlook}
\label{conclusions}
In this work, the formulation of the Newtonian equations describing a collisionless, self-gravitating fluid with velocity-dispersion has been derived both in standard Eulerian coordinates $(\vb x,t)$, as well as in Hubble-comoving coordinates $(\vb q,t)$, leading to the general Euler-Jeans-Newton system under the hypothesis of truncation of the third reduced velocity moment. Then the Lagrangian formulation of this system has been derived, leading to the Lagrange-Jeans-Newton system, consisting of only 10 independent equations (4 field equations and 6 evolution equations for the components of $\Pi_{ij}$) instead of 14 in the Euler-Jeans-Newton system (as Euler and continuity equations are automatically solved in the Lagrangian picture).\footnote{Both of these systems are overdetermined, but can be rewritten (over a simply connected spatial domain) using the gravitational potential $\Phi$ such that $\vb g=-\grad\Phi$ with $\Phi$ the solution of Poisson's equation $\laplacian\Phi=4\pi G\rho-\Lambda$ instead of the gravitational field equations for $\vb g$, $\boldsymbol{\nabla} \times \vb g={\bf 0},\boldsymbol{\nabla} \cdot \vb g=\Lambda-4\pi G\rho$, where the existence of $\Phi$ is ensured by Poincar\'e's lemma.}\smallbreak
Using the Lagrangian framework, the special case of plane symmetry was studied, reducing the four field equations to only one nontrivial equation for the perturbation field $P_1$. This case would describe the collapse of a string embedded into a three-dimensional relativistic background. The remaining equation is nonlinear and does not admit analytical solutions, but it can be linearized and then solved analytically using Fourier analysis (see \cite{al_roumi_matiere_2011}). Trying another path to approximate the solution of the general, nonlinear equation, the ``adhesion approximation'' has been applied to the Euler-Jeans-Newton system, resulting into an equation for the peculiar-velocity field in the form of Burgers' equation, where the effective viscosity coefficient depends on both space and time. From there, two further approximations were made, the first being to merely get rid of the spatial dependence and considering only a time-dependent coefficient, and the second to consider a constant coefficient, the value of which being obtained by time-averaging the previous time-dependent coefficient. The interest of the latter approximation is the existence of rather simple analytical solutions to Burgers' equation with constant coefficient. To compare these two new models to the linearized Lagrangian equation, both the exact equation and the three approximations were integrated numerically and results compared with each other at the time of shell-crossing. Overall, all three approximations are in good agreement with the exact solution, but the ``adhesion approximation'' performs a bit worse than the 
Lagrange-linear approximation. This may come from the neglected spatial dependence of the viscosity coefficient, which can be taken into account in the form of Equation (\ref{burgers+}). It might be interesting to investigate numerical solutions to this equation in order to extend the idea of the ``adhesion approximation''.\smallbreak
Investigating on the intuitive idea that the size and density of the initial collapsing cloud has an impact on the dynamics of the said collapse, the exact Lagrangian equation was integrated for various initial domain sizes and densities, and the time of first shell-crossing was identified. Results show that the crossing time expressed in terms of the scale factor $a(t)-1$ is, as expected, a decreasing function of the initial density and an increasing function of the initial size of the cloud. For very small or very large clouds, this behavior is well-approximated by a power law of the initial cloud size, and overall exhibits a nontrivial dependence on the dimensionless parameter $\tilde\beta$ linked to the initial density. Further investigation of the parameter space is needed to conclude on any analytical relation between those quantities.\smallbreak
In the last section of this work, the Lagrangian formalism was applied to the general case of a three-dimensional collapsing cloud without any particular symmetry. In this case it has been shown that the dynamics of the collapse can be rendered \textit{locally} one-dimensional using physical considerations on the magnitude of the invariants of the perturbation tensor. Within this approximation, rather simple evolution equations were found for the various kinematic parts of the Lagrangian perturbation tensor. In particular, the equation for its trace part is similar to the one obtained in the plane-symmetric approximation, except for the equation of state linking the trace of the velocity dispersion tensor to the density. An interesting further work would be to study this equation under various approximations. In particular, dropping the nonlinear term and assuming a spatially constant proportionality between $P$ and $\Delta_{ij|ji}$ (which can be justified by the coinciding eigendirections of the tensors) may open possible analytical developments on the obtained master equation. 

\bigskip
\noindent
{\bf Acknowledgements:}
This work is part of a project that has received funding from the European Research Council (ERC) under the European Union's Horizon 2020 research and innovation program (grant agreement ERC advanced grant 740021--ARTHUS, PI: TB). We thank the anonymous referee for insightful and constructive comments that helped to improve the paper.

\newpage

\begin{appendix}

\section{Proofs}
\label{appA}
\setcounter{equation}{0}
\renewcommand{\theequation}{\thesection.\arabic{equation}}
\subsection{Moments of the Vlasov equation}
We consider the phase space density as a function of Eulerian phase space coordinates and time, $f(x_i,v_i,t)$, and we start by specifying our integration domain. We take a domain $\Omega^v$ included in the velocity phase subspace. We first calculate the zeroth velocity moment of the Vlasov equation, expressing the conservation of the phase space volume \cite{binney_galactic_2008}:
\begin{equation}
    m\int_{\Omega^v}\qty(\pdv{f}{t}+v_i\pdv{f}{x_i}+g_i\pdv{f}{v_i})\,d^3v.
\end{equation}
The first two terms are readily integrated by swapping the integral and the differential operators:
\begin{gather}
    m\int_{\Omega^v}\pdv{f}{t}=\pdv{t}\int_{\Omega^v} mf\,d^3v=:\pdv{t}\rho,\\
    m\int_{\Omega^v}v_i\pdv{f}{x_i}=\pdv{x_i}\int_{\Omega^v}mv_if\,d^3v=:\pdv{x_i}(\rho\overline{v}_i).
\end{gather}
In the third term, as $\vb g$ does not depend on the velocities, we can take it out of the integral and use Stokes' theorem to write
\begin{equation}
    m\int_{\Omega^v}g_i\pdv{f}{v_i}=mg_i\int_{\del\Omega^v} f\,d^2\!A^v.
\end{equation}
Now we make the physical assumption that the high-velocity states are rare, 
\begin{equation}
 \forall n\in\N\quad v^nf(\cdot,\cdot,\vb v)\xrightarrow[v\to\infty]{}0.
\end{equation}
In practice, we suppose that the phase space density is cut off at some finite velocity that is contained in the velocity subspace. 
Under this assumption, we have in particular that, for a large enough $\Omega^v$, 
\begin{equation}
    \int_{\del\Omega^v} v^nf\,d^2\!A^v=0,
\end{equation}
and the third term we were calculating vanishes, leaving the desired continuity equation for the rest mass density $\rho$. Now, we consider the first velocity moment:
\begin{equation}
    m\int_{\Omega^v}v_i\qty(\pdv{f}{t}+v_j\pdv{f}{x_j}+g_j\pdv{f}{v_j})\,d^3v.
\end{equation}
As before, the first two terms are readily integrated
\begin{gather}
    m\int_{\Omega^v}v_i\pdv{f}{t}=\pdv{t}\int_{\Omega^v} mv_if\,d^3v=:\pdv{t}(\rho\overline{v_i}),\\
    m\int_{\Omega^v}v_iv_j\pdv{f}{x_j}=\pdv{x_j}\int_{\Omega^v}mv_iv_jf\,d^3v:=\pdv{x_j}(\rho\overline{v_iv_j}).
\end{gather}
The third term can be integrated by parts:
\begin{eqnarray}
    &m\int_{\Omega^v}v_i g_j\pdv{f}{v_j}\,d^3v =mg_j\int_{\Omega^v}v_i\pdv{f}{v_j}\,d^3v\nonumber\\
    &=mg_j\qty\bigg(\int_{\del\Omega^v}v_i f\,d^2\!A^v-\int_{\Omega^v} f\underbrace{\pdv{v_i}{v_j}}_{=\delta_{ij}}\,d^3v).
\end{eqnarray}
The integrated term vanishes as $f$ is cut off, and we are left with
\begin{equation}
    m\int_{\Omega^v}v_ig_j\pdv{f}{v_j}\,d^3v =-g_i\int_{\Omega^v}mf\,d^3v=-\rho g_i,
\end{equation}
which yields the momentum conservation equation. To obtain the Euler-Jeans equation, we start by injecting the Lagrangian time-derivative into this equation, which yields:
\begin{equation}
    \dv{t}(\rho\overline{v}_i)-\overline{v}_j\pdv{x_j}(\rho\overline v_i)=\rho g_i-\pdv{x_j}(\rho\overline{v_iv_j}).
\end{equation}
Now, using the continuity equation,
\begin{equation}
    \dv{t}\rho=-\rho\boldsymbol{\nabla}\cdot\overline{\vb v},
\end{equation}
we rewrite
\begin{equation}
    \dv{t}(\rho\overline{v}_i)=\rho\dv{t}\overline v_i+\overline{v}_i\dv{t}\rho=\rho\dv{t}\overline v_i-\rho\overline v_i\pdv{x_j}\overline v_j,
\end{equation}
which leads to 
\begin{gather}
    \rho\dv{t}\overline v_i=\rho g_i-\pdv{x_j}(\rho\overline{v_iv_j})+\overline{v}_j\pdv{x_j}(\rho\overline v_i)+\rho\overline v_i\pdv{x_j}\overline v_j\nonumber\\
    =\rho g_i-\pdv{x_j}(\rho\overline{v_iv_j})+\pdv{x_j}\qty(\rho\overline v_i\overline v_j)\nonumber\\
    =\rho g_i-\pdv{x_j}\qty(\rho\qty(\overline{v_iv_j}-\overline v_i\overline v_j)) =: \rho g_i- \pdv{x_j}\Pi_{ij},
\end{gather}
where we have defined the reduced second velocity moment $\Pi_{ij} = \rho \overline{(v_i - \overline v_i)(v_j - \overline v_j)}=\rho\qty(\overline{v_iv_j}-\overline v_i\overline v_j)$.
\subsection{Evolution of the reduced second velocity moment}
We calculate the second velocity moment of the Vlasov equation, using the same techniques as above, and find
\begin{equation}
    \pdv{t}\qty\big(\rho\overline{v_iv_j})+\pdv{x_k}\qty\big(\rho\overline{v_iv_jv_k})=\rho\qty\big(g_i\overline v_j+g_j\overline v_i).
\end{equation}
Re-expressing the third reduced velocity moment, 
\begin{eqnarray}
L_{ijk} : = \rho \overline{(v_i - \overline v_i)(v_j - \overline v_j)(v_k - \overline v_k)} \nonumber\\
=\rho\qty(\overline{v_iv_jv_k}-\overline v_i\overline v_j \overline v_k)-\qty[\overline v_i\Pi_{jk}+\overline v_j\Pi_{ki}+\overline v_k\Pi_{ij}],\ 
\end{eqnarray}
writing the Lagrangian time-derivative of $\Pi_{ij}$, 
\begin{equation}
\dv{t}\Pi_{ij} =  \pdv{t}\qty\big(\rho\overline{v_iv_j})-\pdv{t}\qty\big(\rho\overline v_i\overline v_j)+\overline v_k 
  \pdv{x_k}\Pi_{ij},
\end{equation}
and using the previous velocity moment equations, we arrive at
\begin{equation}
  \dv{t}\Pi_{ij}=-\qty[\pdv{\overline v_k}{x_k}\Pi_{ij}+\pdv{\overline v_i}{x_k}\Pi_{jk}+\pdv{\overline v_j}{x_k}\Pi_{ki}]-\pdv{x_k}L_{ijk}.
\end{equation}
\subsection{EJN system for peculiar-quantities}
To show Equations (\ref{ejn_pec}), we recall the expressions of the homogeneous quantities: ${\vb v}_H=\dot a\vb q,\vb g_H=\ddot a\vb q, \rho_H=\rho_{H_0}/a^3$, and $\pi^H_{ij}=p_Ha^2\delta_{ij}$. Injecting the decomposition (\ref{peculiar}) into (\ref{ejna}), one obtains: 
\begin{equation}
    \dv{t}\qty[(1+\delta)\rho_H]+(1+\delta)\rho_H\qty(\pdv{{ v_H}_k}{x_k}+\pdv{u_k}{x_k})=0.
\end{equation}
The Lagrangian time-derivative is invariant under the transformation from inertial Eulerian to Hubble-comoving Eulerian coordinates, so that we only have to transform the Eulerian derivatives. Expanding the different terms, we rewrite the above expression in the form
\begin{gather}
    \rho_H\dv{t}\delta+\dfrac{1}{a}(1+\delta)\rho_H\pdv{u_k}{q_k}+(1+\delta)\qty[\dv{t}\rho_H+\rho_H\dfrac{1}{a}\pdv{{v_H}_k}{q_k}]\nonumber\\=0.
\end{gather}
We now show that the bracketed term is zero, using the expressions for $\rho_H$ and ${\vb v}_H$:
\begin{gather}
    \dv{t}\rho_H+\rho_H\dfrac{1}{a}\pdv{ { v_H}_k}{q_k}=\dv{t}\dfrac{\rho_{H_0}}{a^3}+\dfrac{\rho_{H_0}}{a^3}\dfrac{\dot a}{a}\pdv{q_k}{q_k}\nonumber\\
    =-3\rho_{H_0}\dfrac{\dot a}{a^4}+3\rho_{H_0}\dfrac{\dot a}{a^4}=0.
\end{gather}
This leaves us with
\begin{equation}
    \dv{t}\delta+\dfrac{1}{a}(1+\delta)\pdv{u_k}{q_k}=0.
\end{equation}
Now, we move to the momentum conservation equation:
\begin{gather}
    (1+\delta)\rho_H\dv{t}\qty({v}_{H_i}+u_i)\nonumber\\
    =(1+\delta)\rho_H({g_H}_i+w_i)-\pdv{x_j}\qty(\pi^H_{ij}+\pi_{ij}).
\end{gather}
As before, we transform the equation into the Hubble-comoving coordinates, and we make use of the expressions for the background quantities. In particular, as $\pi^H_{ij}$ is homogeneous, its gradient vanishes, leaving
\begin{gather}
    (1+\delta)\rho_H\dv{t}\qty(\dot a q_i)+(1+\delta)\rho_H\dv{t}u_i\nonumber\\
    =(1+\delta)\rho_H\ddot a q_i+(1+\delta)\rho_H w_i-\dfrac{1}{a}\pdv{q_j}\pi_{ij}.
\end{gather}
As an aside, we calculate
\begin{equation}
    \dv{t}(\dot aq_i)=\ddot a q_i+\dot a\dot q_i=\ddot a q_i+Hu_i,
\end{equation}
which gives after reinjecting in the above equation 
\begin{gather}
    (1+\delta)\rho_H(\ddot a q_i+Hu_i)+(1+\delta)\rho_H\dv{t}u_i\nonumber\\
    =(1+\delta)\rho_H\ddot aq_i+(1+\delta)\rho_Hw_i-\dfrac{1}{a}\pdv{q_j}\pi_{ij},
\end{gather}
which simplifies to
\begin{equation}
    \dv{t}u_i+Hu_i=w_i-\dfrac{1}{a}\dfrac{1}{(1+\delta)\rho_H}\pdv{q_j}\pi_{ij}.
\end{equation}
We continue with the evolution equation for the reduced peculiar-velocity moment tensor $\pi_{ij}$:
\begin{eqnarray}
    &\dv{t}\qty(p_Ha^2\delta_{ij}+\pi_{ij})\nonumber\\
    &=-\dfrac{1}{a}\Bigg[\qty(\dot a\pdv{q_i}{q_k}+\pdv{u_i}{q_k})\qty(p_Ha^2\delta_{jk}
    +\pi_{jk})\nonumber\\
    &+\qty(\dot a\pdv{q_j}{q_k}+\pdv{u_j}{q_k})\qty(p_Ha^2\delta_{ki}+\pi_{ki})\nonumber\\
    &+\qty(\dot a\pdv{q_k}{q_k}+\pdv{u_k}{q_k})\qty(p_Ha^2\delta_{ij}+\pi_{ij})\Bigg].
\end{eqnarray}
First, we can expand the l.h.s and use the identity $\del_{q_j}q_i=\delta_{ij}$ on the r.h.s, then expand it:
\begin{gather}
    2p_H\dot aa\delta_{ij}+\dv{t}\pi_{ij}\nonumber\\
    = -\dfrac{1}{a}\bigg[\qty(\dot a\delta_{ik}+\pdv{u_i}{q_k})\qty(p_Ha^2\delta_{jk}+\pi_{jk})\nonumber\\
    +\qty(\dot a\delta_{jk}+\pdv{u_j}{q_k})\qty(p_Ha^2\delta_{ki}+\pi_{ki})\nonumber\\
    +\qty(3\dot a+\pdv{u_k}{q_k})\qty(p_Ha^2\delta_{ij}+\pi_{ij})\bigg]\nonumber\\
    -\dfrac{1}{a}\bigg[5p_H\dot aa^2\delta_{ij}+p_Ha^2\qty(\pdv{u_i}{q_j}+\pdv{u_j}{q_i}+\pdv{u_k}{q_k}\delta_{ij})+5\dot a\pi_{ij}\nonumber\\
    +\pdv{u_i}{q_k}\pi_{jk}+\pdv{u_j}{q_k}\pi_{ki}+\pdv{u_k}{q_k}\pi_{ij}\bigg].
\end{gather}
Finally, regrouping similar terms and rearranging the expression, we obtain:
\begin{gather}
    \dv{t}\pi_{ij}+5H\pi_{ij}\nonumber\\
    =-\dfrac{1}{a}\qty(\pdv{u_i}{q_k}\pi_{jk}+\pdv{u_j}{q_k}\pi_{ki}+\pdv{u_k}{q_k}\pi_{ij})\nonumber\\
    -p_Ha^2\qty[\dfrac{1}{a}\qty(\pdv{u_i}{q_j}+\pdv{u_j}{q_i})+\qty(\dfrac{1}{a}\pdv{u_k}{q_k}+7H)\delta_{ij}].
\end{gather}
The irrotationality constraint on the gravitational field becomes
\begin{equation}
    \dfrac{\ddot a}{a}\underbrace{\grad_{\vb q}\times\vb q}_{=\vb 0}+\dfrac{1}{a}\grad_{\vb q}\times\vb w={\bf 0}\Rightarrow\grad_{\vb q}\times\vb w={\bf 0}, 
\end{equation}
while the divergence equation becomes
\begin{gather}
    \dfrac{\ddot a}{a}\pdv{q_k}{q_k}+\dfrac{1}{a}\pdv{w_k}{q_k}=\Lambda-4\pi G\rho_H(1+\delta)\nonumber\\\Leftrightarrow \  3\dfrac{\ddot a}{a}+\dfrac{1}{a}\pdv{w_k}{q_k}=\Lambda-4\pi G\rho_H(1+\delta),
\end{gather}
and with the Friedmannian acceleration law for the (eventually relativistic) background,
\begin{equation}
    3\dfrac{\ddot a}{a}=\Lambda-4\pi G\qty(\rho_H+3p_H),
\end{equation}
we finally have
\begin{equation}
    \pdv{w_k}{q_k}=4\pi Ga(3 p_H -\rho_H\delta).
\end{equation}
In the Newtonian case, there is no background pressure, so one may take $p_H=0$. Keeping terms in $p_H$ accounts for the relativistic pressure in the background, which may or may not be relevant. In this paper, we are eventually neglecting it for our purpose of comparing solutions.
\subsection{Lagrangian field equations}
Proving Equations (\ref{field_vec}) to (\ref{lagr_raw}) is straightforward, making use of usual vector calculus identities and of the Lagrangian form of the Eulerian derivative. To prove Equation (\ref{lagr_homog}), we write on the one hand, using Leibniz rule:
\begin{gather}
    \dfrac{1}{2\rho J}\J\qty(\dfrac1J\J(\Pi_{kp},f_q,f_r),f_k,f_i)\nonumber\\
    =\dfrac{1}{2\rho J}\Bigg[\dfrac1J\J(\J(\Pi_{kp},f_q,f_r),f_k,f_i)\nonumber\\
    -\dfrac{1}{J^2}\J(\Pi_{kp},f_q,f_r)\J(J,f_k,f_i)\Bigg]\nonumber\\
    =\dfrac{1}{2C_HJ}\J(\J(\Pi_{kp},f_q,f_r),f_k,f_i)\nonumber\\
    -\dfrac{1}{2C_HJ^2}\J(\Pi_{kp},f_q,f_r)\J(J,f_k,f_i),
\end{gather}
writing $\rho=C_H/J$ with ${C_H}_{|i}=0$. On the other hand,
\begin{gather}
    -\dfrac{1}{2\rho^2J^2}\J(\rho,f_k,f_i)\J(\Pi_{kp},f_q,f_r)\nonumber\\
    =-\dfrac{1}{2C_H^2}\J\qty(\dfrac{C_H}{J},f_k,f_i)\J(\Pi_{kp},f_q,f_r)\nonumber\\
    =\dfrac{1}{2C_HJ^2}\J\qty(J,f_k,f_i)\J(\Pi_{kp},f_q,f_r),
\end{gather}
so that these terms cancel out with the second term of the first expression, leaving only the desired terms in (\ref{lagr_homog}). To prove (\ref{lagr_p}) from there, we expand $\Pi_{kp}=p\delta_{kp}+\Delta_{kp}$ and make use of the properties of contracted Levi-Civita tensors, to obtain:
\begin{gather}
    \dfrac{1}{J}\J(\ddot f_k,f_k,f_i)+\dfrac{1}{2C_HJ}\epsilon_{kqr}\J(\J(p,f_q,f_r),f_k,f_i)\nonumber\\+\dfrac{1}{2C_HJ}\epsilon_{pqr}\J(\J(\Delta_{kp},f_q,f_r),f_k,f_i)=0 ;\\
    \dfrac{1}{2J}\epsilon_{kij}\J(\ddot f_k,f_i,f_j)+\dfrac{1}{2C_HJ}\J(\J(p,f_i,f_j),f_i,f_j)\nonumber\\
    +\dfrac{1}{4C_HJ}\epsilon_{kij}\epsilon_{pqr}\J(\J(\Delta_{kp},f_q,f_r),f_i,f_j)=\Lambda-4\pi G\rho.
\end{gather}
The first and last terms of each equation are already those that appear in (\ref{lagr_p}), so that we are only interested in the terms involving $p$. The one in the first equation can be shown to be zero. Indeed, using the equation of state $p=\alpha(\rho)$, 
\begin{gather}
    \dfrac{1}{2C_HJ}\epsilon_{kqr}\J(\J(p,f_q,f_r),f_k,f_i)\nonumber\\
    =\dfrac{1}{2C_HJ}\epsilon_{kqr}\J(\alpha'\J(\rho,f_q,f_r),f_k,f_i)\nonumber\\
    =\dfrac{1}{2C_HJ}\epsilon_{kqr}\Bigg[\alpha'\J(\J(\rho,f_q,f_r),f_k,f_i)\nonumber\\
    +\alpha''\J(\rho,f_q,f_r)\J(\rho,f_k,f_i)\Bigg],
\end{gather}
and by antisymmetry properties, the last term in the brackets is zero. For the first term, we rewrite
\begin{gather}
    \dfrac{\alpha'}{2C_HJ}\epsilon_{kqr}\J(\J(\rho,f_q,f_r),f_k,f_i)\nonumber\\
    =\dfrac{\alpha'}{2J}\epsilon_{kqr}\J\qty(\J\qty(\dfrac{1}{J},f_q,f_r),f_k,f_i)\nonumber\\
    =\dfrac{\alpha'}{2J}\epsilon_{kqr}\Bigg[-\dfrac{1}{J^2}\J(\J(J,f_q,f_r),f_k,f_i)\nonumber\\+\dfrac{2}{J^3}\J(J,f_q,f_r)\J(J,f_k,f_i)\Bigg].
\end{gather}
Here, we can again use the antisymmetry properties to get rid of the last term, but we may actually keep a part of it, to put the remaining terms into the form: 
\begin{gather}
    \dfrac{\alpha'}{2C_HJ}\epsilon_{kqr}\J(\J(\rho,f_q,f_r),f_k,f_i)\nonumber\\
    =-\dfrac{\alpha'}{J}\epsilon_{kqr}\Bigg[\dfrac{1}{2J^2}\J(\J(J,f_q,f_r),f_k,f_i)\nonumber\\
    -\dfrac{1}{2J^3}\J(J,f_q,f_r)\J(J,f_k,f_i)\Bigg],
\end{gather}
where the bracketed term is the Lagrangian form of the vector identity $\boldsymbol{\nabla}\times \grad J={\bf 0}$, hence the first equation of (\ref{lagr_p}) is proven. \smallbreak\noindent
For the second equation, the proof is straightforward:
\begin{gather}
    \dfrac{1}{2C_HJ}\J(\J(p,f_i,f_j),f_i,f_j)\nonumber\\
    =\dfrac{1}{2C_HJ}\Bigg[\alpha'\J(\J(\rho,f_i,f_j),f_i,f_j)\nonumber\\+\alpha''\J(\rho,f_i,f_j)\J(\rho,f_i,f_j)\Bigg]\nonumber\\
    =\dfrac{1}{2C_HJ}\Bigg[\alpha'\J\qty(\J\qty(\dfrac{C_H}{J},f_i,f_j),f_i,f_j)\nonumber\\+\alpha''\J\qty(\dfrac{C_H}{J},f_i,f_j)\J\qty(\dfrac{C_H}{J},f_i,f_j)\Bigg]\nonumber\\
    \times\dfrac{1}{2C_HJ}\Bigg[C\alpha'\Bigg(-\dfrac{1}{J^2}\J(\J(J,f_i,f_j),f_i,f_j)\nonumber\\+\dfrac{2}{J^3}\J(J,f_i,f_j)\J(J,f_i,f_j)\Bigg)\nonumber\\
    +\dfrac{\alpha''C_H^2}{J^4}\J(J,f_i,f_j)\J(J,f_i,f_j)\bigg]\nonumber
\end{gather}
\begin{gather}      
    =\qty(\dfrac{C_H\alpha''}{2J^5}+\dfrac{\alpha'}{J^4})\J(J,f_i,f_j)\J(J,f_i,f_j)\nonumber\\-\dfrac{\alpha'}{2J^3}\J(\J(J,f_i,f_j),f_i,f_j).
\end{gather}
\subsection{1D isotropic pressure}
We wish to prove the stated solution of Equation (\ref{pi11_dot}). We postulate a solution of the form $\pi_{11}=\beta a^\mu\rho^\gamma$, with $\beta$ an arbitrary constant. We then have
\begin{gather}
    \pdv{\pi_{11}}{t}=\beta a^\mu\rho^{\gamma-1}\qty(\mu\rho H+\gamma\pdv{\rho}{t})\\
    \pdv{\pi_{11}}{q_1}=\beta\gamma a^\mu\rho^{\gamma-1}\pdv{\rho}{q_1}.
\end{gather}
We also recall the continuity equation
\begin{gather}
    \pdv{\rho}{t}+\dfrac{u_1}{a}\pdv{\rho}{q_1}
    =-\dfrac{\rho}{a}\grad_{\vb q}\cdot\qty(\dot a\vb q+\vb u)=-3\rho H-\dfrac{\rho}{a}\pdv{u_1}{q_1}.
\end{gather}
Injecting everything into Equation (\ref{pi11_dot}), we obtain:
\begin{equation}
    (5+\mu-3\gamma)\rho H+(3-\gamma)\dfrac{\rho}{a}\pdv{u_1}{q_1}=0,
\end{equation}
which is generally satisfied only for $\gamma=3$ and $\mu=4$.
\subsection{Expressions for $a$ and $b$}
We start by rewriting Friedmann's equation (restricted to vanishing background pressure $p_H$) in the adimensional form,
\begin{equation}
    \ddot a=-\dfrac{\Omega_m^0}{2a^2}+\Omega_\Lambda^0a,
\end{equation}
where $\Omega_m^0=\tfrac{8\pi G\rho_{H,0}}{3H_0^2}$ and $\Omega_\Lambda^0=\tfrac{\Lambda}{3H_0^2}$ are the matter and dark energy density parameters at the present time in the FLRW universe model, and the overdot now denotes the derivative with respect to $\tau$. One can easily verify that the given expression for $a(\tau)$ satisfies this equation. We then obtain $h$ by differentiation. Now, starting from the equation for $b$ in (\ref{parallel}), and using $a$ as a new time-variable (differentiation with respect to $a$ is denoted by a $'$), we can rewrite
\begin{equation}
    \dot a^2b''+\qty(\ddot a+2\dfrac{\dot a^2}{a})\;b'-\dfrac{3\Omega_m^0}{2a^3}b=0.
\end{equation}
Integrating Friedmann's equation in the absence of curvature allows to write $\dot a^2$ as a function of $a$,
\begin{equation}
    \dot a^2=\dfrac{\Omega_m^0}{a}+\Omega_\Lambda^0a^2,
\end{equation}
which can then be injected, along with Friedmann's equation, into the previous equation for $b$, yielding
\begin{equation}
    b''+\dfrac{3(r_0+2a^3)}{2a(r_0+a^3)}b'-\dfrac{3r_0}{2a^2(r_0+a^3)}b=0,\label{ba}
\end{equation}
where we introduced $r_0=\Omega_m^0/\Omega_\Lambda^0$. This equation admits non-Liouvillian solutions that can be written in terms of the hypergeometric function $_2F_1$. The change of dependent variable 
\begin{gather}
    y=b\exp\;\left(\int\dfrac{3(r_0+2a^3)}{4a(r_0+a^3)}\dd a\right)\nonumber\\=\qty\big[a^3(a^3+r_0)]^{1/4}b(a)=\qty[\dfrac{r^2x}{(1-x)^2}]^{1/4}b(x),
\end{gather}
maps the above equation to its so-called ``normal form'',
\begin{equation}
    y''=I_1(x)y\qc I_1(x)=\dfrac{3(4a^6+20a^3r+7r^2)}{16a^2(a^3+r)^2},
\end{equation}
where $I_1(x)$ is called the \textit{invariant} of the equation. Two equations such as (\ref{ba}) can be mapped onto each other if their normal form coincides after a change of independent variable $\xi=a^k$. To simplify the calculations, we look for the exponent $k$ such that the exponents found in the numerator of the invariant of the transformed equation, $I_0(\xi)$, are the lowest integers possible. In our case, we make the transformation $\xi=a^3$, resulting in 
\begin{equation}
    I_0(\xi)=-\dfrac{20\xi^2+4r_0\xi+11r_0^2}{144\xi^2(\xi+r_0)^2}.
\end{equation}
The hypergeometric equation 
\begin{equation}
    x(x-1)y''+\qty[(\mu+\nu+1)x-\sigma]y'+\mu\nu y=0,
\end{equation}
admits the normal form (also known as Q-form)
\begin{equation}
    \dfrac{y''}{y}=\dfrac{\qty[(\mu-\nu)^2-1]x^2+2\qty[2\mu\nu-(\mu+\nu-1)\sigma]x+\sigma(\sigma-2)}{4x^2(x-1)^2}.
\end{equation}
We now look for a M\"obius' transformation of the independent variable
\begin{equation}
    x=\dfrac{\alpha\xi+\beta}{\gamma\xi+\delta},
\end{equation}
mapping the invariant of the hypergeometric equation, $I_{_2F_1}(x)$ to $I_0(\xi)$. After the transformation, one can write
\begin{equation}
    I_{_2F_1}(\xi)=\dfrac{\omega_2\xi^2+2\omega_1\xi+\omega_0}{(\sigma_1\xi+\sigma_2)^2(\sigma_3\xi+\sigma_4)^2(\sigma_5\xi+\sigma_6)^2},
\end{equation}
where
\begin{subequations}
\begin{equation}
    \omega_0=\Delta^2\qty(\beta^2A+2\beta\delta B+\delta^2 C)
\end{equation}\vspace{-.9cm}
\begin{equation}
    \omega_1=\Delta^2\qty(\alpha\beta A+\qty(\alpha\delta+\beta\gamma)B+\gamma\delta C)
\end{equation}\vspace{-.9cm}
\begin{equation}
    \omega_2=\Delta^2\qty(\alpha^2A+2\alpha\gamma B+\gamma^2 C)
\end{equation}\vspace{-.9cm}
\begin{equation}
    \sigma_1=\alpha\qc\sigma_2=\beta
\end{equation}\vspace{-.9cm}
\begin{equation}
    \sigma_3=\gamma\qc\sigma_4=\delta
\end{equation}\vspace{-.9cm}
\begin{equation}
    \sigma_5=\alpha-\gamma\qc\sigma_6=\beta-\delta
\end{equation}\vspace{-.9cm}
\begin{equation}
    \Delta=\alpha\delta-\beta\gamma
\end{equation}\vspace{-.9cm}
\begin{equation}
    A=\dfrac{1}{4}\qty\big[(\mu-\nu)^2-1]
\end{equation}\vspace{-.9cm}
\begin{equation}
    B=\dfrac{1}{4}\qty[2\mu\nu-(\mu+\nu-1)\sigma]
\end{equation}\vspace{-.9cm}
\begin{equation}
    C=\dfrac{1}{4}\sigma(\sigma-2) \ .
\end{equation}
\end{subequations}
We first identify the poles of both invariants, which yields the coefficients of the M\"obius transform:
\begin{equation}
    x=\dfrac{\xi}{\xi+r_0}.
\end{equation}
Then we inject these values in the coefficients $\omega_i$ of the numerator, and we find the possible hypergeometric coefficients:
\begin{equation}
    \mu=\dfrac{5}{6}\qc \nu=\dfrac{1}{3}\qc \sigma=\dfrac{11}{6}.
\end{equation}
The solutions of this hypergeometric equation then yield the solution of (\ref{ba}) by reverting all the transformations. Solutions satisfying the correct boundary condition $b(a\to0)\sim a$ and $\lim_{a\to\infty}b<\infty$ yield both growing and decaying modes:
\begin{subequations}
\begin{gather}
    b^-=(r_0x)^{-1/2}\\
    b^+=\dfrac{5}{6}\dfrac{r_0^{1/3}}{\sqrt{x}}B_x\qty(\dfrac{5}{6},\dfrac{2}{3}).
\end{gather}
\end{subequations}
\section{Numerical methods}
\label{appB}
\setcounter{equation}{0}
\subsection{Adimensional equations}
In all the following, the overdot denotes the derivative with respect to the rescaled time variable $\tau=H_{\rm i}t$.
\subsubsection{1D master equation}
In order to adimensionalize Equation (\ref{p1_beta}), we define the following rescaled quantities:
\begin{equation}
    \tilde P=\dfrac{P_1}{L_J^{\rm i}}\qc \tilde X=\dfrac{X_1}{L_J^{\rm i}}\qc \tilde\beta=\dfrac{\beta}{\beta_0}\qc L_J^{\rm i}=\sqrt{\dfrac{3\beta_0\rho_{H,\rm i}}{4\pi G}}.
\end{equation}
The scaling factor $L_J^{\rm i}$ is the Jeans length of the system at initial time. Although $\beta$ is a constant, it is not dimensionless and is a priori unknown, giving a degree of freedom in the global magnitude of the viscosity. To emphasize this arbitrary choice, we split $\beta$ into its dimensional value $\beta_0$ and a dimensionless coefficient $\hat\beta$.~In these new variables, Equation~(\ref{p1_beta}) reads:
\begin{equation}
    \pdv[2]{\tilde P}{\tau}+2h\pdv{\tilde P}{\tau}-\dfrac{\sigma}{a^3}\tilde P=-\tilde\beta\dfrac{\sigma}{a^4}\pdv[2]{\tilde P}{\tilde X}\qty(1+\pdv{\tilde P}{\tilde X})^{-4},
\end{equation}
where $\sigma=3\Omega_m^{\rm i}/2$. The parameters $a$ and $\sigma$ are determined by the initial time $\tau_{\,\rm i}$, and the only free parameter is $\tilde\beta$. This equation is of second order in time, and although it might be integrated numerically in its current form, it is easier and more adapted to our solver (see Appendix B.2) to rewrite it as a system of two first order PDEs in time. Let 
\begin{equation}
    \xi=\tilde P\qc \eta=\pdv{\tilde P}{\tau}, 
\end{equation}
then the previous equation can be written as 
\begin{equation}
    \left\{\begin{array}{l}
        \dot\xi=\eta  \\
        \dot\eta=-2h\eta+\dfrac{\sigma}{a^3}\xi-\tilde\beta\dfrac{\sigma}{a^4}\dfrac{\xi''}{(1+\xi')^4},
    \end{array}\right.
\end{equation}
where an overdot denotes a partial derivative with respect to time and a prime denotes a partial derivative with respect to space. Note that the linearized version of this system can be written in the matrix form
\begin{equation}
\renewcommand{\arraystretch}{1.3}
    \dot Z=\begin{bmatrix}
    0&1\\
    \dfrac{\sigma}{a^3}\qty(1-\tilde\beta\dfrac{\sigma}{a}\displaystyle\pdv[2]{\tilde X})&-2h
    \end{bmatrix}Z\qc Z=\begin{bmatrix}\xi\\\eta\end{bmatrix}.
\end{equation}
\subsubsection{Burgers' equation}
In order to adimensionalize Equation (\ref{burgers}), we use the previous rescaled variables, to which we add
\begin{equation}
    \hat u=\dfrac{\tilde u}{L_J^{\rm i}}.
\end{equation}
In these new variables, Burgers' equation reads:
\begin{equation}\label{adim_burgers}
    \pdv{\hat u}{b}+\hat u\pdv{\hat u}{\hat q}=\hat\mu\pdv[2]{\hat u}{\hat q}\qc \hat\mu=\hat\beta\dfrac{3\Omega_m^{\rm i}}{2}(1+\delta)\dfrac{b}{a^4\dot b}.
\end{equation}
The dependence in $\delta$ will be put aside in the numerical integration of this equation.
\subsection{Numerical scheme}
Consider a generic partial differential equation (PDE) of the form
\begin{equation}
    \pdv{u}{t}=L(u)+N(u)+g,
\end{equation}
where $u$ is the unknown function of space and time, $L$ is a linear spatial differential operator acting on $u$, $N$ is a nonlinear operator acting on $u$ (that does not imply any time-differentiation) and $g$ is a given function of space and time. Let $\Delta t, \Delta x$ be the time and space steps and $N,J$ be the number of time and space intervals, so that the discretized space and time variable read $\qty{x_j}=x_0,\dots,x_{J}$ and $\qty{t_n}=t_0,\dots,t_N$. Let $\vb U^n$ be the approximation of the solution at time $t_n$. We denote respectively by $\mathbb{L}^n$, $\vb N(\vb U^n)$ and $\vb g^n$ the matrix representation of the linear operator $L$, the approximation of $N(u)$ and of $g$ at time $t_n$, after taking into account boundary conditions. In all the calculations, the first spatial derivatives will be computed using a first-order centered differences scheme and the second spatial derivatives with the standard three-point centered differences scheme:
\begin{equation}
    \pdv{u}{x}\longrightarrow\dfrac{U_{j+1}^n-U_{j-1}^n}{2\Delta x}\qc\pdv[2]{u}{x}\longrightarrow\dfrac{U_{j+1}^n-2U_j^n+U_{j-1}^n}{\Delta x^2}.
\end{equation}
At each time step, the evolution of $u$ is approximated using a second-order accurate Crank-Nicholson scheme for the linear part of the equation, and a second-order accurate Adams-Bashforth scheme for the nonlinear part. Taking into account the periodic boundary conditions that sum up to identifying $x_0$ with $x_J$, we obtain the following expression at each time step:
\begin{subequations}
\begin{gather}
    U^{n+1}_J=U^{n+1}_0,\\
    \vb U^{n+1}_{[0:J-1]}=\Bigg(\mathds 1-\dfrac{\Delta t}{2}\mathbb L^{n+1}\Bigg)^{-1}\ \mathfrak{O}\nonumber\\
    \mathfrak{O}=\Bigg[\Bigg(\mathds 1+\dfrac{\Delta t}{2}\mathbb L^{n}\Bigg)\vb U^{n}_{[0:J-1]}
    +\dfrac{\Delta t}{2}\Bigg(3\vb N\Bigg(\vb U^{n}_{[0:J-1]}\Bigg) \nonumber\\
    -\vb N\Bigg(\vb U^{n-1}_{[0:J-1]}\Bigg)\Bigg)
    +\dfrac{\Delta t}{2}\Bigg(\vb g^{n+1} + \vb g^n \Bigg)\Bigg],
\end{gather}
\end{subequations}
where $\vb U^{n}_{[0:J-1]}$ denotes the vector constituted of the first $J$ components of $\vb U^n$. The first time step cannot use a second order Adams-Bashforth scheme for the nonlinear term, and we therefore jump-start the solver using a simple forward Euler scheme:
\begin{subequations}
\begin{gather}
    U^{1}_J=U^{1}_0,\\
    \vb U^{1}_{[0:J-1]}=\qty(\mathds 1-\dfrac{\Delta t}{2}\mathbb L^{1})^{-1}\Bigg[\qty(\mathds 1+\dfrac{\Delta t}{2}\mathbb L^{0})\vb U^{0}_{[0:J-1]}\nonumber\\
    +\Delta t\,\vb N\qty(\vb U^{0}_{[0:J-1]})+\dfrac{\Delta t}{2}\qty(\vb g^{1} + \vb g^0)\Bigg].
\end{gather}
\end{subequations}
As an example, let us consider Burgers' equation (\ref{adim_burgers}), which reads (letting all the hats down for the sake of readability):
\begin{equation}
    \pdv{u}{b}=\mu\pdv[2]{u}{q}-u\pdv{u}{q}.
\end{equation}
Here, there is no source term hence $g=0$. We identify the linear and nonlinear terms,
\begin{equation}
    L=\mu\pdv[2]{q}\qc N(u)=-u\pdv{u}{q},
\end{equation}
and their following discrete approximation, taking into account periodic boundary conditions, reads:
\begin{equation}
\renewcommand{\arraystretch}{1.3}
    \mathbb{L}^n=\dfrac{\mu(t_n)}{\Delta q^2}\begin{bmatrix}
        -2 & 1 & & & 1 \\
        1 &  &\ddots &  & \\
         & \ddots & \ddots & \ddots & \\
         & &\ddots & & 1\\
        1 & & & 1 & -2 
    \end{bmatrix}_{J\times J};
    \end{equation}
    \begin{equation}
    \vb N\qty(\vb U^n_{[0:J-1]})=\dfrac{1}{2\Delta q}\begin{bmatrix}
        U_0^n\qty(U_{J-1}^n-U_1^n) \\
        U_1^n\qty(U_0^n-U_2^n) \\
        \vdots \\
        U_{J-2}^n\qty(U_{J-3}^n-U_{J-1}^n) \\
        U_{J-1}^n\qty(U_{J-2}^n-U_{0}^n) 
    \end{bmatrix}.
\end{equation}
\end{appendix}
\newcommand\eprintarXiv[1]{\href{http://arXiv.org/abs/#1}{arXiv:#1}}


\begin{thebibliography}{50}           
%
\bibitem{planck_collaboration}
P.A.R. Ade {\it et al.} Planck 2015 results. XIII. Cosmological parameters.
Astron. Astrophys. \href{https://doi.org/10.1051/0004-6361/201525830}{{\bf 594}, A13 (2016)}.
[\eprintarXiv{1502.01589}]
%
\bibitem{adler_lagrangian_1998}
S. Adler and T. Buchert, 
Lagrangian theory of structure formation in pressure-supported cosmological fluids.
Astron. Astrophys. \href{https://doi.org/}{\textbf{343}, 317 (1999)}
[\eprintarXiv{astro-ph/9806320}]
%
\bibitem{al_roumi_matiere_2011}
F. Al Roumi, Mati\`ere sombre: R\'etroaction et analogie champ scalaire, in French, {\it Internship report},
	\'Ecole Normale Sup\'erieure de Lyon, (2011).
%
\bibitem{rza4}
F. Al Roumi, T. Buchert and A. Wiegand,
Lagrangian theory of structure formation in relativistic cosmology. IV. Lagrangian approach to gravitational waves.
Phys. Rev. D \href{https://doi.org/10.1103/PhysRevD.96.123538}{\textbf{96}, 123538 (2017)}. 
[\eprintarXiv{1711.01597}]
%
\bibitem{rza3}
A. Alles, T. Buchert, F. Al Roumi and A. Wiegand,
Lagrangian theory of structure formation in relativistic cosmology. III. Gravitoelectric perturbation and solution schemes at any order.
Phys. Rev. D \href{https://doi.org/10.1103/PhysRevD.92.023512}{{\bf 92}, 023512 (2015)}
[\eprintarXiv{1503.02566}]
%
\bibitem{bildhaueretal}
S. Bildhauer, T. Buchert and M. Kasai,
Solutions in Newtonian cosmology -- The pancake theory with cosmological constant.
Astron. Astrophys. {\bf 263}, 23 (1992).
%
\bibitem{binney_galactic_2008}
J. Binney and S. Tremaine, Galactic dynamics.
{\it Princeton Series in Astrophysics}
\href{https://press.princeton.edu/books/paperback/9780691130279/galactic-dynamics}{ISBN},
Princeton University Press (2008).
%
\bibitem{bonnor_jeans_1957}
W.B. Bonnor, 
Jeans' formula for gravitational instability.
M.N.R.A.S. \href{https://doi.org/10.1093/mnras/117.1.104}{\textbf{117}, 104 (1957)}.
%
\bibitem{buchert_varenna}
T. Buchert, Lagrangian perturbation approach to the formation of large-scale structure.
in {\it Proc. International School Enrico Fermi} 
Course CXXXII {\it Dark Matter in the Universe}, Varenna 1995, ed. S. Bonometto, 
J.R. Primack and A. Provenzale, IOS Press Amsterdam, pp.543-64 (1996).
[\eprintarXiv{astro-ph/9509005}]
%
\bibitem{nonperturbative}
T. Buchert, The non-perturbative regime of cosmic structure formation.
Astron. Astrophys. \href{https://doi.org/10.1051/0004-6361:20064899}{{\bf 454}, 415 (2006)}.
[\eprintarXiv{astro-ph/0601513}]
%
\bibitem{correspondence}
T. Buchert, A direct correspondence between Newtonian gravitation and general relativity.
Phys. Rev. Lett., under review.
[\eprintarXiv{2306.12253}]
%
\bibitem{buchert:adhesive}
T. Buchert and A. Dom\'\i nguez,
Modeling multi-stream flow in collisionless matter: Approximations for large-scale structure beyond shell-crossing.
Astron. Astrophys. {\bf 335}, 395 (1998).
[\eprintarXiv{astro-ph/9702139}]
%
\bibitem{buchert_adhesive_2005}
T. Buchert and A. Dom\'\i nguez,
Adhesive gravitational clustering.
Astron. Astrophys. \href{https://doi.org/10.1051/0004-6361:20052885}{
{\bf 438}, 443 (2005)}.
[\eprintarXiv{astro-ph/0502318}]
%
\bibitem{extending}
T. Buchert, A. Dom\'\i nguez and J. P\'erez-Mercader, 
Extending the scope of models for large-scale structure formation in the Universe.
Astron. Astrophys. {\bf 349}, 343 (1999).
[\eprintarXiv{astro-ph/9709218}]
%
\bibitem{buchert_geometrical_2009}
T. Buchert, G.F.R. Ellis and H. van Elst,
Geometrical order-of-magnitude estimates for spatial curvature in realistic models of the Universe.
Gen. Rel. Grav. \href{https://doi.org/10.1007/s10714-009-0828-4}{\textbf{41}, 2017 (2009)}. 
%
\bibitem{bks}
T. Buchert, M. Kerscher and C. Sicka, 
Backreaction of inhomogeneities on the expansion: The evolution of cosmological parameters.
Phys. Rev. D 
\href{https://doi.org/10.1103/PhysRevD.62.043525}{{\bf 62}, 043525 (2000)}.
[\eprintarXiv{astro-ph/9912347}]
%
\bibitem{rza1}
T. Buchert and M. Ostermann, 
Lagrangian theory of structure formation in relativistic cosmology. I. Lagrangian framework and definition of a nonperturbative approximation.
Phys. Rev. D \href{https://doi.org/10.1103/PhysRevD.86.023520}{\textbf{86} 023520 (2012)}.
[\eprintarXiv{1203.6263}]
%
\bibitem{Universe}
T. Buchert, I. Delgado Gaspar and J.J. Ostrowski,
On general-relativistic Lagrangian perturbation theory and its nonperturbative generalization.
Universe \href{https://doi.org/10.3390/universe8110583}{\textbf{8} 583 (2022)}. [\eprintarXiv{2209.13417}]
%
\bibitem{ruth:vorticity}
G. Cusin, V. Tansella and R. Durrer,
Vorticity generation in the Universe: A perturbative approach.
Phys. Rev. D \href{https://doi.org/10.1103/PhysRevD.95.063527}{\textbf{95} 063527 (2017)}.
[\eprintarXiv{1612.00783}]
%
\bibitem{doroshkevich:vorticity}
A.G. Doroshkevich, 
The Origin of Rotation of Galaxies.
Phys. Rev. Lett. \href{https://ui.adsabs.harvard.edu/abs/1973ApL....14...11D/abstract}{\textbf{14} 11 (1973)}. 
%
\bibitem{doroshkevichetal}
A.G. Doroshkevich, V.S. Ryabenkii and S.F. Shandarin,
Non-linear theory of development of potential perturbations.
Astrofizika {\bf 9}, 257; (1975): Astrophysics \href{https://link.springer.com/article/10.1007/BF01011421}{{\bf 9}, 144 (1975)}.
%
\bibitem{ehlersbuchert}
J. Ehlers and T. Buchert,
Newtonian cosmology in Lagrangian formulation: Foundations and perturbation theory.
Gen. Rel. Grav. \href{https://doi.org/10.1023/A:1018885922682}{{\bf 29}, 733-764 (1997)}.
[\eprintarXiv{astro-ph/9609036}]
%
\bibitem{ehlersrienstra}
J. Ehlers and W. Rienstra,
The locally isotropic solutions of the Liouville and Poisson equations.
The Astrophys. J. \href{https://doi.org/10.1086/149852}{{\bf 155}, 105 (1969)}.
%
\bibitem{gurbatov89}
S.N. Gurbatov, A.I. Saichev and S.F. Shandarin, 
The large-scale structure of the universe in the frame of the model equation of non-linear diffusion.
M.N.R.A.S. \href{https://doi.org/10.1093/mnras/236.2.385}{{\bf 236}, 385 (1989)}.
%
\bibitem{kevorkian}
J. Kevorkian, 
Partial differential equations: analytical solution techniques. 
\href{https://link.springer.com/book/9780534122164}{ISBN}, 
{\it Wadsworth \& Brooks/Cole Math. Series}, Pacific Grove, California (1990).
%
\bibitem{klypinshandarin}
A.A. Klypin and S.F. Shandarin,   
Three-dimensional numerical model of the formation of large-scale structure in the Universe. 
M.N.R.A.S. \href{https://doi.org/10.1093/mnras/204.3.891}{{\bf 204}, 891 (1983)}.
%
\bibitem{lifshitz46}
E.M. Lifshitz, 
On the gravitational stability of the expanding universe.
ZhETF {\bf 16}, 587; J. Phys. USSR Acad. Sci. {\bf 10}, 116 (1946); Republication: Gen. Rel. Grav. \href{https://doi.org/10.1007/s10714-016-2165-8}{{\bf 49}, 18 (2017)}.
%
\bibitem{lifshitzkhalatnikov}
E.M. Lifshitz and I.M. Khalatnikov,  
Investigations in relativistic cosmology.
Usp. Fiz. Nauk {\bf 80},
391 (1963); Sov. Phys. Usp. {\bf 6}, 522 (1964); 
Adv. Phys. \href{https://doi.org/10.1080/00018736300101283}{{\bf 12}, 185 (1963)}.
%
\bibitem{melottshandarin89}
A.L. Melott and S.F. Shandarin, 
Gravitational instability with high resolution.
The Astrophys. J. \href{https://doi.org/10.1086/167681}{{\bf 343}, 26 (1989)}.
%
\bibitem{peebles:book}
P.J.E. Peebles, 
The Large-Scale Structure of the Universe.
\href{https://press.princeton.edu/books/paperback/9780691209838/the-large-scale-structure-of-the-universe}{ISBN}, 
Princeton Univ. Press (1980).
%
\bibitem{rampffrisch}
C. Rampf and U. Frisch, 
Shell-crossing in quasi-one-dimensional flow.
M.N.R.A.S. \href{https://doi.org/10.1093/mnras/stx1613}{\textbf{471}, 671 (2017)}.
[\eprintarXiv{1705.08456}]
%
\bibitem{rampf:shellcrossing1}
C. Rampf and O. Hahn, 
Shell-crossing in a $\Lambda$CDM Universe.
M.N.R.A.S. \href{https://doi.org/10.1093/mnrasl/slaa198}{\textbf{501}, L71 (2021)}.
[\eprintarXiv{2010.12584}]
%
\bibitem{rampf:shellcrossing2}
C. Rampf, U. Frisch and O. Hahn, 
Unveiling the singular dynamics in the cosmic large-scale structure.
M.N.R.A.S. \href{https://doi.org/10.1093/mnrasl/slab053}{\textbf{505}, L90 (2021)}.
[\eprintarXiv{1912.00868}]
%
\bibitem{colombi:shellcrossing1}
S. Saga, A. Taruya  and S. Colombi,
Lagrangian cosmological perturbation theory at shell crossing.
Phys. Rev. Lett. \href{https://doi.org/10.1103/PhysRevLett.121.241302}{\textbf{121} 241302 (2018)}.
[\eprintarXiv{1805.08787}]
%
\bibitem{colombi:shellcrossing2}
S. Saga, A. Taruya and S. Colombi,
Cold dark matter protohalo structure around collapse: Lagrangian cosmological perturbation theory versus Vlasov simulations.
Astron. Astrophys. \href{https://doi.org/10.1051/0004-6361/202142756}{\textbf{664} A3 (2022)}.
[\eprintarXiv{2111.08836}]
%
\bibitem{serrin}
I. Serrin, 
Mathematical Principles of Classical Fluid Mechanics.
In: {\it Encycl. of Phys.} \href{https://doi.org/10.1007/978-3-642-45914-6_2}{{\bf VIII.1} 125--263}, Springer (1959).
%
\bibitem{shukurov}
A.M. Shukurov, 
Nonlinear growth of density perturbations in an expanding collisionless medium.
Astrofizika {\bf 17}, 469; Astrophysics  \href{https://doi.org/10.1007/BF01005587}{{\bf 17}, 263 (1982)}. 
%
\bibitem{vigneron:darkmatter}
Q. Vigneron and T. Buchert, 
Dark matter from backreaction? Collapse models on galaxy cluster scales. 
Class. Quantum Grav. \href{https://doi.org/10.1088/1361-6382/ab32d1}{\textbf{36} 175006 (2019)}.
[\eprintarXiv{1902.08441}]
%
\bibitem{zeldovich70a}
Ya.B. Zel'dovich,   
Gravitational instability: an approximate theory for large density perturbations.
{\it Astron. Astrophys.} {\bf 5}, 84 (1970).
%
\bibitem{zeldovich70b}
Ya.B. Zel'dovich,   
Fragmentation of a homogeneous medium under the action of gravitation.
{\it Astrofizika} {\bf 6}, 319; {\it Astrophysics} \href{https://doi.org/10.1007/BF01007263}{{\bf 6}, 164 (1973)}.
%
\end{thebibliography}
\end{document}